\documentclass[twocolumn,nofootinbib]{aastex6}
\usepackage{graphicx}
\usepackage{amssymb}
\usepackage{multirow}
\usepackage{amsmath}
\usepackage{mathptmx}
\usepackage{booktabs}
\usepackage{float}

\newcommand\degree{$^{\circ}$}

\newcommand\RBNS{{\mathcal{R}_{\rm BNS}}}
\newcommand{\msun}{\,{\rm M_{\odot}}}
\newcommand{\MBH}{\,{M_{{\rm BH}}}}

\newcommand{\Epmaxprime}{\,{E'_{p,\rm max}}}
\newcommand{\tdecayi}{\,{t_{{\rm decay},i}}}
\newcommand{\tdecayp}{\,{t_{{\rm decay},\pi}}}
\newcommand{\tdecayk}{\,{t_{{\rm decay},K^+}}}

\newcommand{\Ekisoprime}{\,{E'_{k,\rm iso}}}

\newcommand{\SU}{\,{\it SU1}}
\newcommand{\SUa}{\,{\it SU2}}
\newcommand{\SUb}{\,{\it SU3}}
\newcommand{\CU}{\,{\it CU}}
\newcommand{\WM}{\,{\it WM}}
\newcommand{\CW}{\,{\it CW}}
\newcommand{\HM}{\,{\it HM}}
\newcommand{\tpg}{\,{t_{p\gamma}}}
\newcommand{\tpgi}{\,{t_{p\gamma,i}}}

\newcommand{\tc}{\,{t_{{\rm cool}}}}
\newcommand{\tci}{\,{t_{{\rm cool},i}}}
\newcommand{\tcp}{\,{t_{{\rm cool},\pi}}}

\newcommand{\fsupi}{\,{f_{\rm sup,i}}}
\newcommand{\fsupp}{\,{f_{\rm sup,\pi}}}
\newcommand{\fsupk}{\,{f_{\rm sup,K^+}}}

\newcommand{\Egm}{\,{E_{\gamma,{\rm min}}}}
\newcommand{\thobs}{\,{\theta_{\rm obs}}}
\newcommand{\cm}{\,{{\rm cm}}}

\newcommand\simlt{\lower.5ex\hbox{$\; \buildrel < \over \sim \;$}}
\newcommand\simgt{\lower.5ex\hbox{$\; \buildrel > \over \sim \;$}}

\usepackage{lineno}

\begin{document}
\title{The role of jet-cocoon mixing, magnetization and shock breakout in neutrino and cosmic-ray emission from short GRBs}
\author{Ore Gottlieb\altaffilmark{1,2} and No\'emie Globus\altaffilmark{3,4,5}}
\altaffiltext{1}{School of Physics \& Astronomy, Tel Aviv University, Tel Aviv 69978, Israel. E-mail: ore@northwestern.edu}
\altaffiltext{2}{Center for Interdisciplinary Exploration \& Research in Astrophysics (CIERA), Physics \& Astronomy, Northwestern University, Evanston, IL 60208, USA}
\altaffiltext{3}{ELI Beamlines, Institute of Physics, Czech Academy of Sciences, 25241 Dolní Břežany, Czech Republic. E-mail:  globus@fzu.cz}
\altaffiltext{4}{Center for Computational Astrophysics, Flatiron Institute, Simons Foundation, New-York, NY10003, USA}
\altaffiltext{5}{Astrophysical Big Bang Laboratory (ABBL), Cluster for Pioneering Research, RIKEN, 2‐1 Hirosawa, Wakō, Saitama 351‐0198, Japan}

\begin{abstract}
We perform GRMHD and RMHD simulations of weakly and highly magnetized gamma-ray burst (GRB) jets propagating in binary neutron star (BNS) merger ejecta. Using the simulations, we first find that mixing between the jet and cocoon, which is present in all types of jets, inhibits the formation of subphotospheric collisionless shocks. However, we show that a mild magnetization may lead to the formation of collisionless subshocks which allow efficient proton acceleration. We consider shear acceleration and diffusive shock acceleration at collimation shocks, internal shocks, shock breakout, and external shocks, to provide the first estimate for neutrino and cosmic-ray (CR) signals from self-consistent simulations of GRBs in BNS mergers. We find that short GRBs do not produce detectable neutrino signals with current-day facilities. Shock breakout yields $ \sim 10 $ PeV neutrinos at viewing angles $\sim20^\circ $, independent of the jet magnetization. However, a neutrino signal from shock breakout is well below detection limits of current detectors. Such a signal would allow a coincident neutrino-$\gamma-$ray detection, providing a testable prediction for shock breakout as a neutrino production site. Using the numerical modeling that fits GW170817 afterglow emission, we find that blast waves in BNS mergers can account for 5\%-10\% of the Galactic CR luminosity in the PeV-EeV energy range. Based on these estimates, the observed level of CR anisotropy places a constraint on the distance of the latest Galactic binary neutron star merger to $\lesssim3$ kiloparsecs. 
\end{abstract}
\section{Introduction}
\small
The year 2017 was a breakthrough in multi-messenger astronomy, with the detection of the binary neutron star (BNS) merger GW170817 through gravitational waves (GW) in advanced LIGO/Virgo \citep{Abbott_2017} and electromagnetic emission \citep[see e.g.,][for review]{Nakar2020,Margutti2020}.
However, although $\gamma$-ray bursts (GRB) are considered to be a promising source of neutrinos \citep[e.g.,][]{Waxman1997}, no neutrino associated with the event has been detected by IceCube \citep{albert2017},
presumably due to the large viewing angle with respect to the jet axis, $ \thobs = {20^\circ}^{+8^\circ}_{-7^\circ}$ \citep{Mooley2018b}. 
Future neutrino detectors will enable us to connect neutrino and electromagnetic signals.
With the majority of BNS mergers with a detectable signal of jets  expected at viewing angles outside the jet opening angle \citep{Gottlieb2019a}, understanding the dependency of the neutrino flux upon the viewing angle is of utmost importance.

The GW170817's multiwavelength campaign showed evidence for a structured jet, as the origin of the $\gamma$-ray emission and the first months of X-ray and radio emission originated in a mildly relativistic ($\Gamma \sim 3$) cocoon  \citep[e.g.,][]{Gottlieb2018b}. The cocoon, which is inflated during the interplay between the jet and the merger's ejecta, applies pressure on the jet and could form a collimation shock at the jet's base.
The jet may successfully break out of the ejecta or be choked inside. Regardless of the jet's fate, the cocoon eventually breaks out of the ejecta, radiating its internal energy upon breakout, in what is likely to be the origin of GRB170817A \citep[][]{Gottlieb2018b}. 
The question then arises if there is a specific signature in energetic particles associated with the fate of the jet that is connected to the nature of the central engine, which is still under debate. One possible jet production mechanism is the magnetic extraction of the spin energy of a Kerr black hole \citep[BZ,][]{blandford1977}. If the inner regions of the disk are dominated by neutrino cooling, the large amount of MeV neutrinos annihilate in the polar region to produce a hot, pressure-driven ${\rm e}^{+}{\rm e}^{-}$ outflow that may interfere with the magnetic energy extraction \citep{globus2014}. Therefore, the jet could be either weakly or highly magnetized, depending on the plasma injection processes near the black hole (BH) horizon. As the launching mechanism is still under debate, unmagnetized, mildly and highly magnetized jets should all be investigated for neutrino production.

Previous studies considered neutrino emission in GRB jets at their collimation and internal shocks \citep[e.g.,][]{Murase2013,globus2015grb,Biehl2018,Kimura2018a}, and shear turbulence between the jet and the cocoon \citep[e.g.,][]{rieger2004,sahayanathan2009,Rieger2016}.
Additional studies also examined neutrino emission from sources outside the jet such as supernova shock breakout \citep[e.g.,][]{Waxman2001,Katz2011,Kashiyama2013} and cocoons of successful or choked jets \citep[e.g.,][]{Ostrowski2000,Senno2016,He2018}.
These studies estimated analytically the structure of the shock, energy reservoir and magnetic dissipation at the aforementioned production sites. However, the jet dynamics involves a non-linear evolution of the complex jet-cocoon system which makes relativistic magnetohydrodynamic simulations an essential tool for studying the jet evolution in BNS mergers \citep[e.g.,][]{Christie2019,Fernandez2019,Kathirgamaraju2019}. One particularly important property that emerges in numerical simulations is the mixing between the jet and the cocoon. In unmagnetized and weakly magnetized jets, the mixing comes about following the development of hydrodynamic instabilities on the jet-cocoon boundary, loading the jet with baryons and reducing its Lorentz factor \citep[e.g.,][]{Gottlieb2021b,Gottlieb2021a,Matsumoto2021}.
The mixing, which has been ignored in previous works, loads the jet with baryons, and thus inhibits the formation of collisionless shocks which are necessary for neutrino production.

BZ-launched GRB jets have a large degree of magnetization which has a tremendous effect on the neutrino production. First, the magnetization significantly alters the jet structure such that the energy is distributed differently between the jet and the cocoon \citep{Gottlieb2020b}. Second, and more importantly, the magnetization needs to be neither too high nor too low for neutrino production below the photosphere.
On one hand, the jet needs to be sufficiently magnetized to support the formation of subphotospheric collisionless shocks. Upon launching, the jet has to initially punch through the optically-thick ejecta, forming radiation-mediated shocks. In such shocks the scattering mean free path is much larger than the Larmor radius implying that such shocks are inefficient for particle acceleration \citep[see e.g.,][]{Levinson2008}. If the jet is magnetized, strong subshocks may emerge within the radiation-mediated shock, on a Larmor radius scale \citep[][hereafter B17]{{Beloborodov2017}}.
On the other hand, in the presence of very strong magnetic fields, pions, which constitute the main neutrino production channel, cool down by synchrotron much faster than their half-life time, thereby significantly mitigating the neutrino production \citep[see e.g.,][]{Lipari2007}. Therefore, the diffusive shock acceleration (DSA) efficiency and hence, the neutrino outcome are strongly dependent upon the evolution of the magnetization, which can be studied with numerical simulations.

CR production from both mildly and ultra relativistic DSA has  been studied in GRB jets following their interaction with the surrounding medium in the afterglow phase \citep[][]{Vietri1995,Dermer2006}. 
In the context of BNS mergers, it has been shown that  Galactic BNS (GBNS) remnants may significantly contribute to the observed CR in the energy range between the knee and the ankle, the so called "shin" region \citep{Kimura2018b}.
However, despite the valuable information about the blast wave structure and energetics that was provided by GW170817 afterglow, the contribution of the jet-cocoon blast-wave to CR remained unexplored.

In this {\it Letter}, we investigate particle production in short GRBs (sGRBs) by performing 3D relativistic magneto-hydrodynamic (RMHD) and general relativistic magneto-hydrodynamic (GRMHD) simulations of structured jets propagating in a BNS ejecta, using  GW170817 observations for calibrating the energetics of the jet-cocoon system. We consider neutrino production by DSA in jets with different magnetizations in regions where strong collisionless subshocks can form. That is: collimation shocks, internal shocks and breakout of the forward shock from the ejecta. We also examine shear acceleration between the jet and the cocoon. Finally, we consider DSA at the external relativistic blast-wave to estimate CR production.

In \S\ref{sec:magnetization}, we show that previous criteria for the formation of collisionless shocks in the jet are not satisfied in numerical simulations, and magnetic fields are necessary for the formation of collisionless subshocks.
In \S\ref{sec:simulations}, we briefly describe the setup of our simulations and where collisionless shocks may form.
In \S\ref{sec:neutrinos} and \S\ref{sec:CR} we present the first neutrino and CR calculations that are based on realistic sGRB simulations, estimated at a variety of production sites and for different types of jets. In \S\ref{sec:summary} and \S\ref{sec:discussion} we summarize and discuss the implications of our results.

\section{Criterion for efficient CR acceleration in radiation-mediated shocks}\label{sec:magnetization}

Subphotospheric shocks are leading candidates for neutrino production by virtue of Fermi acceleration. These shocks are mediated by photons in the optically thick ejecta which supports $ p\gamma $ interactions between the trapped photons and the accelerated protons.
However, efficient Fermi acceleration can only take place in collisionless shocks where particle interactions are mediated by plasma processes, raising the question of whether subphotospheric shocks can efficiently accelerate CR.
\citet{Murase2013} suggested that a collisionless shock can form inside the optically thick region
if the shock's comoving upstream medium is optically thin, namely $ n\sigma_T R' \lesssim 1 $, where $ n $ is the proper number density, $ \sigma_T $ is the Thomson cross section, and $ R' $ is the proper size of the shock.
This criterion can be applied to both oblique collimation shocks and internal shocks which reside inside the optically thick medium.
According to this criterion, \citet{Murase2013} found that neutrino production is disfavored in high-luminosity choked jets, leaving low-luminosity jets as potential neutrino sources.
However, due to the mixing between the jet and the cocoon material, which is present in all types of jets, we find that the above inequality is not satisfied in any type of sGRB jets, whether it is high or low luminosity \citep{Gottlieb2021a}, choked or successful. Thus, collisionless shocks and a neutrino signal are unlikely to emerge under this criterion in sGRBs, as we discuss next.

For typical parameters, the criterion reads that an unmagnetized collimation shock forms a collisionless shock if $ L_{{\rm iso},48}R_{s,10}\Gamma_2^{-3} \lesssim 1 $, where $ Q_x $ denotes the value of the quantity $ Q $ in units of $ 10^x $ times its c.g.s. units. The weakest observed sGRB jets maintain $\gamma$-ray luminosity of approximately $ L_{{\rm iso},48} \approx 10 $ \citep[e.g.,][]{Shahmoradi2015}. That implies that in order to obtain collisionless collimation shocks according to the above criterion, either i) the collimation shock reaches $ R_{s,10} \gg 1 $, which is unlikely for typical sGRB durations, particularly in low luminosity jets that likely propagate at subrelativistic velocities inside the ejecta; or ii) the Lorentz factor is of a few hundreds early on, but the mixing between the jet and the dense ejecta (which is necessary for collimation) inhibits such relativistic velocities at small radii. Thus, the hydrodynamic evolution of sGRB jets seems to rule out the possibility of having neutrino emission from collimation shocks of hydrodynamic jets.

We also find that strong internal shocks, which seem to necessitate a variable engine, do not support the formation of collisionless shocks according to the aforementioned criterion. The reason lies in recent results of \citet{Gottlieb2020a} who found intense mixing from the cocoon into the jet during the low power episodes of the jet. The mixing reduces the Lorentz factor and substantially increases the baryon loading in the jet such that the comoving optical depth increases dramatically, for all jets in general, and due to more intense mixing, in low luminosity jets in particular.

If instead, the jet is magnetized, B17 showed that under certain conditions, strong subphotospheric collisionless subshocks may form within the radiation-mediated shock on a Larmor radius scale.
In particular, the collimation and internal shocks that occur below the photosphere may develop such collisionless subshocks and efficiently accelerate CR in the jet.  B17 found that $ \chi \equiv 2\beta_p $ dictates the shock dissipation efficiency as follows ($ \beta_p $ is  the ratio between the thermal to magnetic pressure, $p_t/p_m$). 
For mildly relativistic shocks, if the upstream magnetization is $ 0.1 \lesssim \sigma_u\equiv \frac{B'^2}{4\pi \rho c^2}\lesssim 1 $ ($ B' $ being the plasma proper magnetic field, $ \rho $ is the mass density) and $ \chi \lesssim 2 $, the photons do not have sufficient energy to fully decelerate the upstream to the downstream velocity, subsequently a collisionless subshock forms to satisfy the jump conditions at the shock.
In the case of $ \chi \gg 1 $, the magnetization is too weak to sustain a strong subshock, and for $ \chi \ll 1 $ the magnetization is too high, and the kinetic energy dissipation efficiency is low. 
This criterion implies that collisionless shocks can emerge during the jet evolution inside the optically-thick ejecta if the jet magnetization is $ \chi \sim 1 $, and $ \sigma_u \gtrsim 0.1 $. The latter also promises that the collisionless subshock dominates the energy in the mildly relativistic shock \citep[see e.g., figure 11 in][]{Levinson2019}.

At highly magnetized, relativistic shocks, Fermi acceleration is known to be inefficient. Using PIC simulations, \citet{Sironi2013} found a critical magnetization of $ \sigma \gtrsim 1\%$, above which the particles cannot travel back upstream even when moving at $\sim c $. They showed that at the critical magnetization efficient acceleration can take place only if the upstream Lorentz factor is $ \Gamma_u \lesssim 5 $.
We thus consider only mildly-magnetized and mildly relativistic shocks \citep[however further studies of CR acceleration efficiency in this regime are needed, see e.g.,][]{Crumley2019}.
Assuming the formation of strong subshocks in this regime, we consider a kinetic energy dissipation efficiency of $ \epsilon_d = 0.2 $ \citep[e.g.,][]{Mimica2010,Komissarov2012}.

\section{Numerical simulations overview}\label{sec:simulations}

We calculate the jet structure and the resulting CR and neutrino emission during different phases of the jet evolution, up to the afterglow phase.
We consider a variety of jet parameters, from hydrodynamic to highly magnetized jets, as well as jets which successfully break out from the ejecta and those that are choked inside.
In all simulations, the initial setup includes a homologous, expanding, subrelativistic ($ v < 0.2 $c) cold ejecta that emerges in the aftermath of the merger. The ejecta mass is chosen to be $ M \approx 0.05 \msun $, as inferred from GW170817 observations, and we choose its radial mass density profile to be $ \rho(r) \propto r^{-2} $. In the hydrodynamic models the dense ejecta is embedded in a lighter mildly relativistic homologous tail ejecta that moves at $ v < 0.6 $c.

Our simulations suggest that the early jet evolution can be largely divided into two types of jets: highly magnetized and unmagnetized/weakly magnetized. The setups of both cases is described below. For the reader interested in the technicalities of the simulations, more details are provided in Appendix~\ref{app:simulations}.

\begin{figure*}
		\centering
		\includegraphics[scale=0.107]{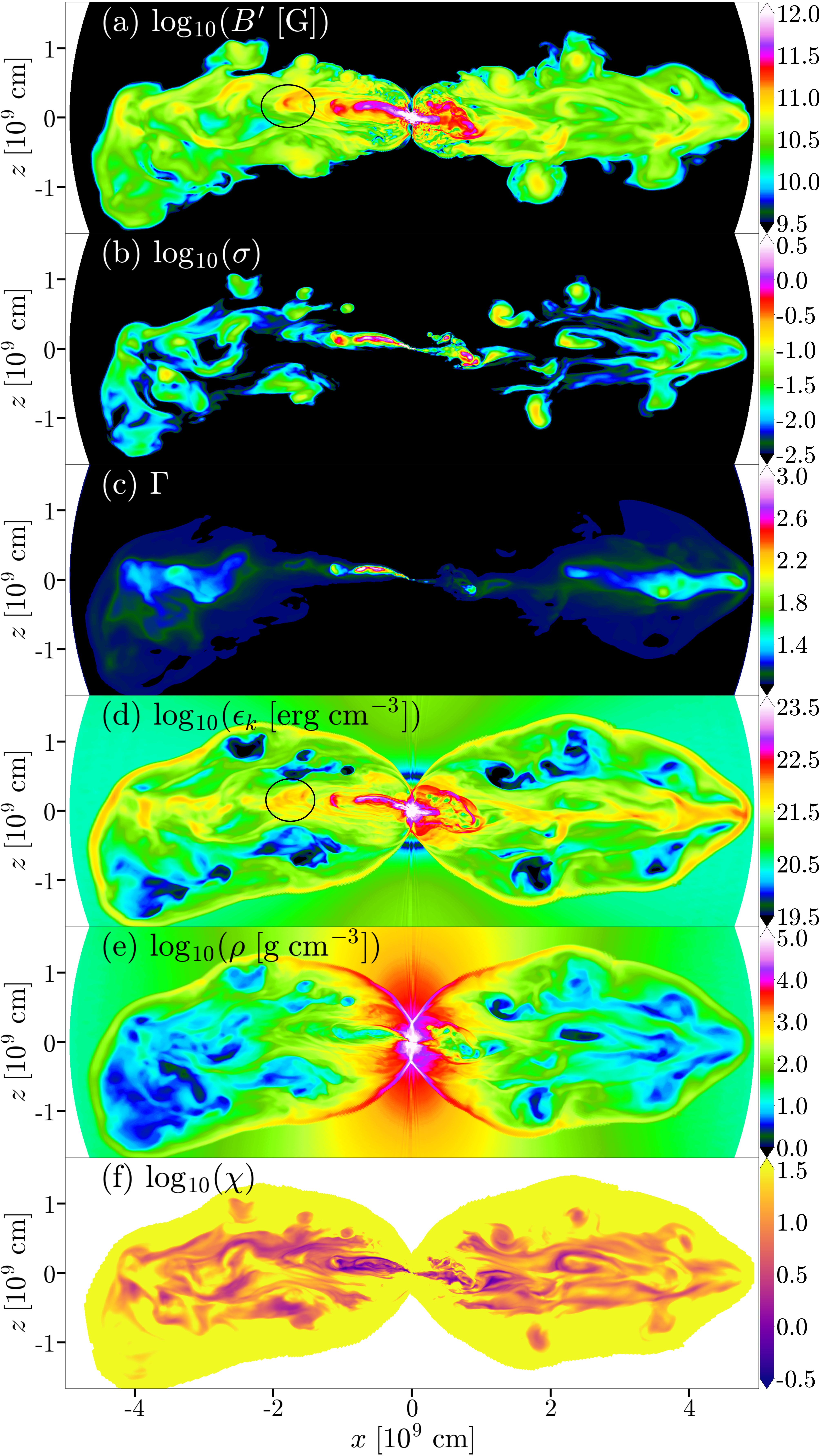}
		\includegraphics[scale=0.31]{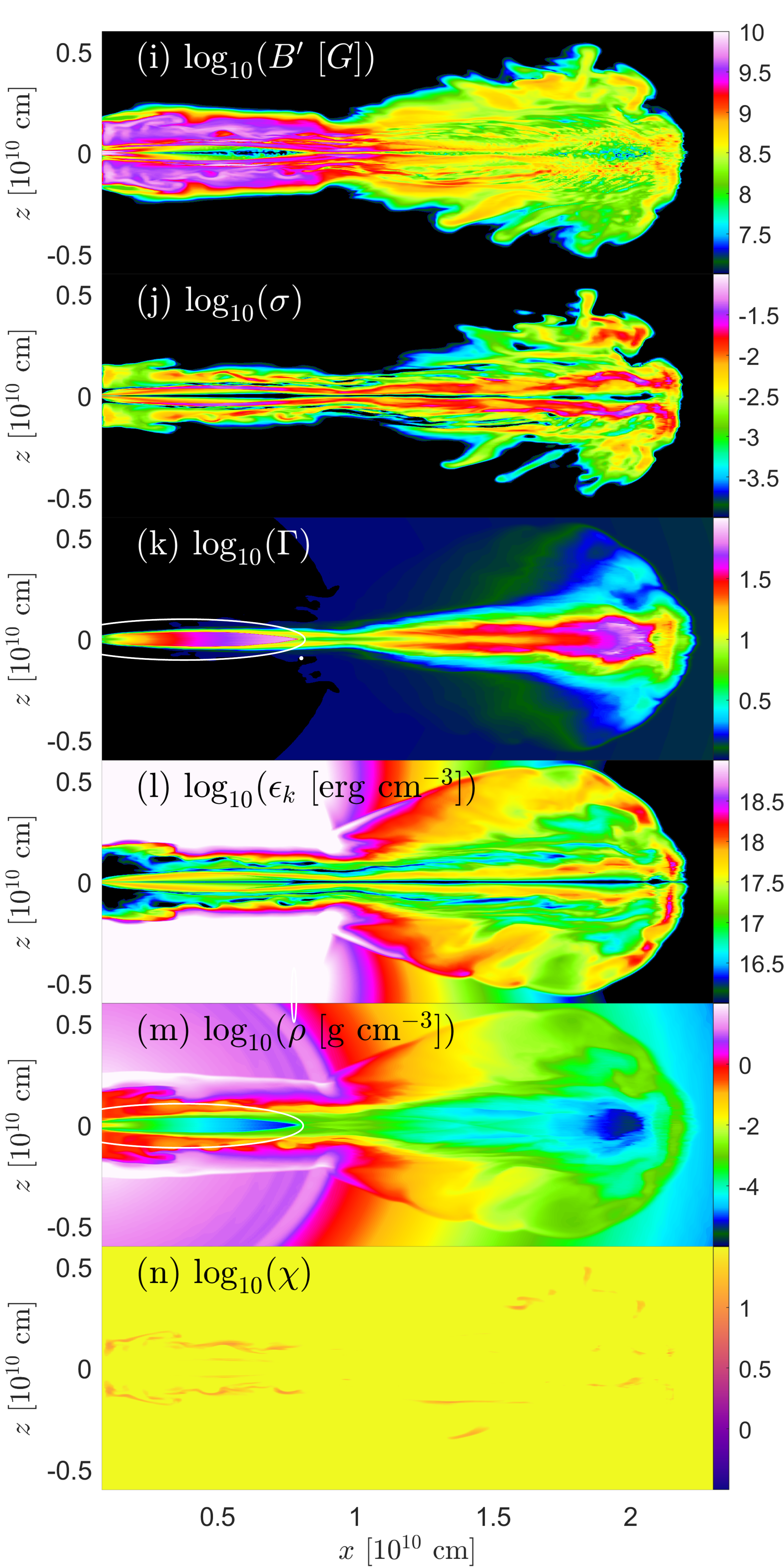}
		\includegraphics[scale=0.32]{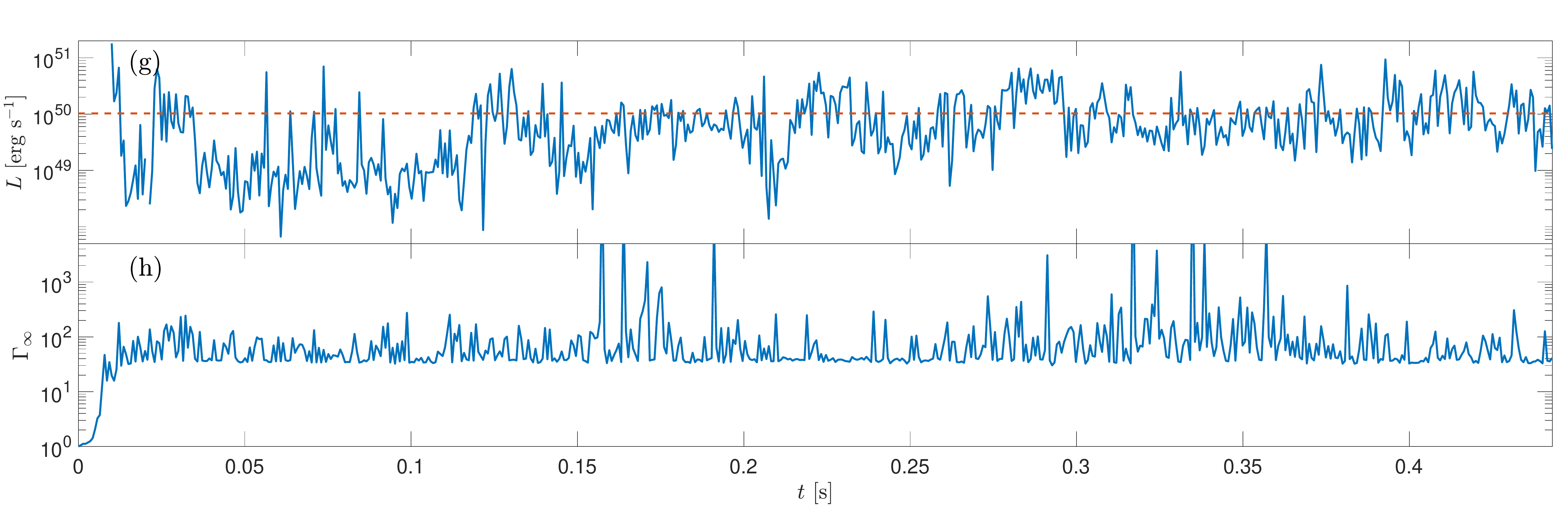}
		\caption[]{Magnetized jets in expanding ejecta. Panels (a-f): Meridional maps of a highly magnetized successful jet (model $ \HM $) inside the ejecta 0.4 s after the jet launch (0.9 s after the merger). Shown are: Logarithmic of the comoving magnetic field (a), magnetization (b), (non logarithmic) Lorentz factor (c), kinetic energy density (d), mass density (e), and $ \chi \equiv 2p_t/p_m $ (f). Black circles mark  examples of internal shocks.
		Panels (g-h): The evolution of the luminosity (g) and the maximal terminal Lorentz factor (product of magnetization and Lorentz factor upon launching, $\sigma_0\Gamma_0$) (h) as measured on the horizon, for the case of highly magnetized jet. Dashed red line in the luminosity (g) panel reflects the average luminosity.
		Panel (i-n): same as panels (a-f) (but Lorentz factor is logarithmic) for a weakly magnetized jet (model $ \WM $) 1s after the jet injection (1.6s after the merger), when the jet already broke out from the ejecta. White ellipses mark the collimation shock.
		In the panels of $ \chi $ we only present the jet and the cocoon, since $ p_t = 0 $ in the unshocked surrounding medium. 
		}
		\label{fig:successful_maps}
\end{figure*}

(i) Highly magnetized (model $ \HM $) jet (performed with the 3D GPU-accelerated general relativistic magneto-hydrodynamic (GRMHD) code \textsc{h-amr} \citep{Liska2019} in Kerr-Schild coordinates, using adaptive mesh refinement (AMR) and local adaptive time-stepping).
The jet launching is affected by the accreted mass, and thus in the GRMHD simulation, where accretion is included, we also consider a more realistic angular distribution\footnote{We also apply this angular dependence in model $ \CU $.}: $ \rho(r,\theta) \propto r^{-2}(\frac{1}{4}+{\rm sin}^3\theta) $ such that most of the mass lies close to the accretion disk on the equatorial plane\footnote{Note that our ejecta is a single component with angular and radial profiles, and the accretion disk forms by the angular momentum of the system. This is in contrast to previous works \citep[e.g.,][]{Christie2019,Fernandez2019} which included the disk as part of their initial conditions.}.
The rest of the initial conditions include a rapidly spinning (dimensionless spin $ a = 0.8 $) Kerr BH with mass $ \MBH = 3\msun $, embedded in the aforementioned expanding ejecta. The ejecta radial density drops linearly from 10 gravitational radii, $ 10 r_g = 10 \,{\cal G}\MBH c^{-2} $ towards $ r_g $. The disk angular velocity is $ 10^{-3}~{\rm s^{-1}} $ for $ r/r_g<70 $, and $ 5(r_g/r)^2~{\rm s^{-1}} $ otherwise. The initial vector-potential profile is a dipole-like with a core of radius $ \sim 100 r_g $, and maximum magnetic field of $ 10^{13} $~G.

The jet is launched self-consistently following the accretion of a magnetized ejecta onto a central BH, and maintains a luminosity of $ L \approx 10^{50} {\rm erg~s^{-1}} $, similar to that in GW170817.
The jet is initially collimated by the winds from the disk, but most of the collimation process is done by the cocoon as can be seen in \url{http://www.astro.tau.ac.il/~ore/NSM_GRMHD.html}).
We follow the jet propagation up to $ 5\times 10^9 $ cm, 0.4 second after its launch.
Fig.~\ref{fig:successful_maps} depicts maps of the jet structure (panels a-f) and its luminosity and maximal terminal Lorentz factor on the event horizon (panels g-h). Movies of the jet evolution can be found in \url{http://www.astro.tau.ac.il/~ore/NSM_GRMHD.html}.
The simulation features a small, highly magnetized collimation shock/nozzle that forms at the base of the jet. The strong magnetic field at the collimation shock of $ B' \approx 10^{12} $ G (panel b) leads to pion and kaon cooling much faster than their decay times, and thus does not support efficient neutrino production.

Our focus in model $ \HM $ is on neutrino production from internal shocks (black circles in Fig. \ref{fig:successful_maps}a,d). We find that within the typical GRB engine time of $ \lesssim 0.5 $ s, internal shocks with the required upstream magnetization $ \sigma_u \gtrsim 0.1 $ and $ \chi \lesssim 2 $ (Fig. \ref{fig:successful_maps}f) primarily take place at $ r \lesssim 3\times 10^9 $ cm where $ \Gamma \lesssim 3 $, thereby being mildly relativistic, and mildly magnetized shocks with $ B' \approx 10^{11} $ G. At later times and larger radii, if the jet engine work-time is sufficiently long, the jet accelerates to ultra-relativistic velocities ($ \Gamma \gtrsim 100 $). At larger radii, the magnetization drops, and thus the shocks become ultra relativistic and weakly-magnetized such that the formation of collisionless shocks is suppressed, and thus we do not consider this phase in our calculation.

(ii) Hydrodynamic (models $ \SU $, $ \SUa $, $ \SUb $) and weakly magnetized (model $ \WM $) jets (performed with the 3D relativistic magneto-hydrodynamic RMHD simulations code \textsc{pluto} \citep{Mignone2007} with constrained transport scheme that keeps $ \nabla \cdot B = 0$).
We choose the jet parameters that are motivated by sGRB observations.
The jets are injected with maximal initial magnetization $ \sigma_0 $ at $ t = t_d $ for a total jet duration of $ t_j $, into the expanding ejecta with a Lorentz factor $ \Gamma_0 \approx 0.7\theta_j^{-1} $ \citep{Mizuta2013,Harrison2018}.
The jet initially accelerates under its thermal pressure to relativistic velocities within an opening angle of $ \theta_j $. As the fluid elements encounter the cocoon, the jet becomes collimated by forming a collimation shock of a size $ R_s $. Following the collimation shock crossing, the jet elements decelerate to mildly relativistic Lorentz factor $ \Gamma \sim \theta_j^{-1} $ \citep[e.g.,][]{Bromberg2011b,Gottlieb2019b}.
The emerging weakly magnetized ($ \WM $) jet-cocoon structure is shown in panels (i-n) in Fig.~\ref{fig:successful_maps}.

Previous numerical simulations of weakly magnetized jets in a dense medium \citep{Gottlieb2020b} suggested that a substantial magnetic dissipation takes place at the collimation shocks, such that the jet magnetization drops to values of $ \sigma \lesssim 0.1 $ following the shocks, disfavoring particle acceleration at internal shocks later on\footnote{Note however that the jet magnetization could be amplified at the collimation and internal shocks to $ \sigma \gtrsim 0.1 $ \citep{Gottlieb2021b}.}. 
When considering the neutrino production at the collimation shock from our weakly magnetized jets (model $ \WM $, white ellipses in Fig.~\ref{fig:successful_maps}k,m), we find that the upstream Lorentz factor is marginally mildly relativistic, but for initial magnetization $ \sigma_0 = 10^{-2} $ (model $ \WM $) the magnetization is too low and $ \chi \gg 1 $ at all potential sites such that no neutrino production is expected prior to breakout.
A mild magnetization of $ \sigma_0 \approx 0.1 $ may allow $ \chi \gtrsim 1 $ at the collimation shock to support neutrino production, however due to numerical limitations we could not perform such a simulation. Instead, we assume in the next section that the collimation shock properties are not considerably different between $ \sigma_0 = 10^{-2} $ and $ \sigma_0 = 10^{-1} $, and consider neutrino production in weakly magnetized jets from their collimation shocks based on our $ \WM $ model.

As the jet-cocoon outflow breaks out of the ejecta the forward shock becomes collisionless, and the magnetization drops substantially such that the outflow can be approximated as hydrodynamic\footnote{However, while the jets are likely to be unmagnetized at late times, the initial magnetization does affect the emerging jet-cocoon structure \citep{Gottlieb2020b}.}. Thus, for the breakout emission and the afterglow phase, we consider successful and choked hydrodynamic jets.
We utilize the fact that the structure of the jet at these stages can be inferred from observations. For the shock breakout emission we average the angular energy and velocity distributions of two models ($ \SUa $ and $ \SUb $), and calibrate their total observed shock breakout energy at $ 20^\circ$ by GRB170817A observations assuming the energy source was shock breakout 
(see Appendix \ref{app:shock_breakout}). For the blast wave emission, we use the jet-cocoon model that fits GW170817 afterglow light curve from \citet{Mooley2018b} (model $ \SU $), and use an additional uncollimated choked jet (model $ \CU $). The jet parameters of all models are shown in \ref{app:simulations}.

\section{Neutrino emission}  \label{sec:neutrinos}

Following the CR acceleration at the collimation shock, internal shocks, forward shock and the shear layer, the CR interact with the background photon field, and generate neutrino emission.
The photon field in mildly relativistic radiation-mediated shocks relaxes to a Wien spectrum as the rest-frame downstream temperature is regulated by pairs to peak at $ \sim 50 $ keV (consistent with the observed peak of $ \sim 100 $ keV), independent of the Lorentz factor of the shock \citep[e.g.,][]{Budnik2010,Nakar2012,Levinson2019}.
We thus assume the photon spectrum to be a Wien spectrum at the collimation shock \citep{Gottlieb2019b} and during the shock breakout \citep{Gottlieb2018b}. Internal shocks may broaden the spectrum to become a Band-like function, and thus for estimating the $ p\gamma $ interaction at the internal shocks, we assume a Band spectrum with $ \alpha_b = -1.2 $ and $ \beta_b = -2.3 $ \citep{Goldstein2013}\footnote{The cooling time of the internal shocks is dominated by synchrotron cooling, thus the specific shape of the spectrum does not significantly affect the results.}. We normalize the photon spectrum based on the available internal energy flux, $ \sim 4p_t c $ at the relevant acceleration site, as found from the simulation (or from GRB170817A for the case of shock breakout). We consider both diffusive shock acceleration in \S\ref{sec:dsa} and shear acceleration in \S\ref{shear}.

\subsection{Diffusive shock acceleration}\label{sec:dsa}
The Fermi process accelerates protons to a power-law distribution $ dN_p/dE_p \propto E_p^{-s} $, where $ s $ is the spectral index. 
We assume a power-law index $ s = 2 $, which roughly holds for mildly relativistic regimes, as it is expected to be $ 2.0 \lesssim s \lesssim 2.2 $ \citep[e.g.,][]{Blandford1987,Keshet2005,Sironi2013}. The rest-frame CR spectrum is:
\begin{equation}\label{eq:CR_spec}
E_p'^2\frac{dN_p}{dE_p'} = \frac{\epsilon_{CR}\epsilon_d\Ekisoprime}{{\rm log}(\Epmaxprime/{\rm GeV})}e^{-\frac{E_p'}{\Epmaxprime}}~,
\end{equation}
where we assumed the minimal CR energy to be the proton rest-mass energy, $ m_pc^2 \approx {\rm GeV} $. The upstream kinetic energy dissipation efficiency is $ \epsilon_d = 0.2 $ in mildly-magnetized shocks (\S\ref{sec:magnetization}) and $ \epsilon_d = 1 $ in shock breakout. We take $ \epsilon_{CR} = 0.1 $ as the fraction deposited in accelerating CR from the rest frame isotropic equivalent dissipated kinetic energy reservoir $ \epsilon_d\Ekisoprime $.

The CR are accelerated by the Fermi process on a comoving timescale,
\begin{equation}
    t'_{acc}(E_p) \approx \frac{E'_p}{\xi_{acc}qB'c} \approx 10^{-6}\frac{E'_{p,12}}{B'_6}~{\rm s}~.
\end{equation}
In the following we use $\xi_{acc}=0.1$ \citep{globus2015grb}.

The maximal energy to which the protons are accelerated $ \Epmaxprime $, is governed by the cooling time of the protons. In general, several cooling processes such as Bethe-Heitler pair production, synchrotron radiation, inverse-Compton radiation, adiabatic cooling, $ pp $ and $ p\gamma $ interactions should be considered. Close to the base of the jet, the high photon number density in the region of the collimation shock renders the photohadron interaction timescale to be the shortest one above the $ p\gamma $ energy threshold. We integrate numerically the $ p\gamma$ cooling time \citep{Berezinskii1990},
\begin{equation}
    \tpg(E_p)=\frac{2\gamma_p^2}{c}\bigg[\int_{\acute{E}_{\gamma,{\rm min}}}^{\infty}\acute{E_\gamma}\kappa_{p\gamma}\sigma_{p\gamma,i} d\acute{E_\gamma}\int_{\frac{\acute{E_\gamma}}{2\gamma_p}}^\infty E_\gamma^{-2}h^{-1}dn'_{\gamma,\nu}dE_\gamma\bigg]^{-1}~,
\end{equation}
where $ h $ is the Planck constant, $ \gamma_p = E_p/m_pc^2 $ is the Lorentz factor of the proton, $ n'_{\gamma,\nu} $ is the downstream spectral photon number density, $ \acute{E}_\gamma $ is the photon energy in the proton frame of reference, $ \kappa_{p\gamma,\pi} $ and $ \sigma_{p\gamma,\pi} $ are the inelasticity and cross section of particle $ i $ in the $ p\gamma $ interaction, respectively\footnote{For the pion cross section and elasticity we use the two-region approximation \citep{Atoyan2003}:
$ \sigma_{p\gamma,\pi} = 340~\mu {\rm b} $, and $ \kappa_{p\gamma} = 0.2 $ when $ \Egm < E_{\gamma} < 3\Egm $, and $ \sigma_{p\gamma,\pi} = 120~\mu {\rm b} $, and $ \kappa_{p\gamma} = 0.6 $ when $ E_{\gamma} > 3\Egm $.
For the kaon cross section we use \citet{Hummer2010}:
$\sigma_{p\gamma,K^+} =
2.0;~3.7;~2.7 ~\mu {\rm b} $ when: $ 6.7\Egm < E_{\gamma} < 8\Egm;~8\Egm < E_{\gamma} < 11\Egm;~11\Egm < E_{\gamma} $, respectively.}.

At large distance from the jet launchpad, where the photon number density drops, synchrotron cooling also plays a role. The synchrotron timescale is given by
\begin{equation}
    t_{syn}(E_p) = \frac{6\pi m_p^4c^3}{m_e^2\sigma_T B'^2 E'_p}~,
\end{equation}
where $ m_e $ is the electron mass.
Finally, as the outflow expands and breaks out, the photon density drops and cooling by proton-proton ($ pp $) interaction may also become important, depending on the ejecta density,
\begin{equation}
    t_{pp} = \frac{m_p}{\sigma_{pp}\kappa_{pp}c\rho_{\rm bo}}~,    
\end{equation}
where $ \rho_{\rm bo} $ is the mass density of the breakout layer, $ \kappa_{pp} \approx 0.5 $ is the proton-proton inelasticity, and $ \sigma_{pp} \approx 5\times 10^{-26}~{\rm cm}^2 $ is the proton-proton cross section at the relevant energies. The total cooling time is $ \tc = (\tpg^{-1}+t_{syn}^{-1}+t_{pp}^{-1})^{-1} $.

The photohadron process becomes important as the $ p\gamma $ cross section increases rapidly for photons with energies above $ E_{\rm th} \approx {1.4\times 10^{17}{\rm eV^2}}/{E_p} $, measured in the proton comoving frame. Above the resonance the interaction yields pions (kaons) which may decay into neutrinos. The efficiency of the process depends on three suppression factors:
i) whether $ p\gamma $ or $ pp $ is the dominant cooling process: $ f_{p\gamma,i} = \frac{\tci}{\tpgi} $, where $ i = \pi (K^+) $ hereafter, and similarly for $ pp $ process, $ f_{pp,\pi} = \frac{\tcp}{t_{pp,\pi}} $;
ii) whether the secondary particle decays before cooling down: $ \fsupi = 1-{\rm exp}\left(-\frac{\tci}{\tdecayi}\right) $, where $ \tdecayp = 2.6\times 10^{-8} $ s and $\tdecayk = 1.2\times 10^{-8} $ s;
iii) the comoving number density of the photon field, $ n_\gamma $: for each proton energy, we estimate the fraction of protons that participate in the $ p\gamma $ interaction by taking the number of the mean free paths along the proper length of the photon target $ R' $, roughly the width of the shock, which is on order of Thomson optical depth \citep[e.g.,][]{Nakar2012}:
\begin{equation}\label{eq:f_pi}
f_{p,i}(E_p) = R'\sigma_{p\gamma,i}\int_{E_{\rm th}}n_{\gamma,\nu}d\nu \approx R'f_\gamma(E_p)n_\gamma\sigma_{p\gamma,i}~,
\end{equation}
where $ \sigma_{p\gamma,i} $ is the $ p\gamma $ cross section for particle $ i $, and $ f_\gamma(E_p) $ is the fraction of the photons that satisfy $ E_\gamma > E_{\rm th} $, which we calculate numerically. If $ f_p(E_p)>1 $, we take $ f_p(E_p)=1$.

In the relevant energy range, $ E \gtrsim 10 $ GeV, the inelasticity of $ p\gamma $ and $ pp $ is $ \kappa \sim 0.5 $, namely half of the protons energy is deposited in the produced pions/kaons. In the main $ p\gamma $ ($pp$) channel, $ f_{p\rightarrow\pi^\pm} = \frac{1}{2} (\frac{2}{3}) $ of the interactions result in charged pions \citep{Anchordoqui2014} that decay to neutrinos via $ \pi^+\rightarrow\mu^++\nu_\mu\rightarrow e^++\nu_e+\nu_\mu+\bar{\nu}_\mu $ and $ \pi^-\rightarrow\mu^-+\bar{\nu}_\mu\rightarrow e^-+\bar{\nu}_e+\nu_\mu+\bar{\nu}_\mu $. The energy is shared rather equally between the four leptons such that $ f_{\pi^\pm\rightarrow\nu_\mu} \approx 0.5 $ of the charged pion energy is transferred to $ \nu_\mu $ and $ f_{\pi^\pm\rightarrow\nu_e} \approx 0.25 $ of it in $ \nu_e $. Averaging over the neutrino flavors and neglecting the weak dependency upon the photon energy,
the maximal rest-frame neutrino energy is $ \kappa f_{p\rightarrow\pi^\pm}(f_{\pi^\pm\rightarrow,\nu_\mu}+f_{\pi^\pm\rightarrow,\nu_e})E_p \approx \frac{3}{16}E_p~(\frac{1}{4}E_p) $ for $ p\gamma $ ($ pp $) interactions, respectively.

When strong magnetic fields are present, pions cool down rapidly and their suppression factor becomes small. Then kaons, whose mass is larger and thus cool slower, become important. The main channel ($ f_{p\rightarrow K^+} = 0.63 $) is $ K^+ \rightarrow \mu^+ + \nu_\mu $, where $ \nu_\mu $ carries $ f_{K^+\rightarrow\nu_\mu} = 0.48 $ of the kaon energy \citep{Lipari2007}. We neglect the energy of the $ \mu^+ $ since it cools down faster than pions, as well as other $ K^+ $ channels as their contribution is at most a few percent,
Overall, the maximal rest-frame energy of neutrinos produced by kaons is $\kappa f_{p\rightarrow K^+}f_{K^+\rightarrow\nu_\mu}E_p \approx \frac{1}{6}E_p $.

The total observed neutrino energy flux is
\begin{equation}
\begin{split}
E_{\nu}^2\frac{dN_{\nu}}{dE_{\nu}} \approx & \bigg[f_{p,\pi}\fsupp\left(\frac{3}{16}f_{p\gamma,\pi}+\frac{1}{4}f_{pp,\pi}\right)+ \\
& f_{p,K^+}\fsupk\frac{\sigma_{p\gamma,K^+}}{\sigma_{p\gamma,\pi}}\frac{1}{6}f_{p\gamma,K^+}\bigg]\delta_D E_p'^2\frac{dN_p}{dE_p'}~,
\end{split}
\end{equation}
where $ \delta_D \equiv [\Gamma(1-\beta{\rm  cos}(\theta-\theta_{\rm obs})]^{-1} $ is the Doppler factor of the shock, $ \beta $ is the dimensionless velocity, $ \theta_{\rm obs} $ is the viewing angle, $ \sigma_{p\gamma,\pi} $, $ \sigma_{p\gamma,K^+} $ are the $ p\gamma $ cross sections for pions and kaons, respectively, $ \fsupi $ is the suppression factor, $ f_{pp,i} $ and $ f_{p\gamma,i} $ are the $ pp $ and $ p\gamma $ dominant cooling factors, respectively, and $ f_{p,i} $ is given by Eq. \ref{eq:f_pi}. 

\subsection{Shear acceleration}\label{shear}
In addition to DSA, another process through which neutrinos can be produced is shear acceleration. In GRBs the most prominent shear region is at the interface between the jet and the cocoon. The shear acceleration time is \citep{rieger2004},
\begin{equation}
   t_{acc}^s = \frac{1}{(4+\zeta)\chi^s\tau} = \frac{\xi_{acc}c}{5\chi^s r_L}~,
\end{equation}
where $\tau$ is the mean scattering time, in its general form $\tau=\tau_0p^\zeta$. In the R.H.S we used Bohm diffusion, $\zeta=1$ and $\tau = r_L/\xi_{acc}c$. $\chi^s$ is the shear coefficient given for a relativistic flow as \citep{rieger2004}:
\begin{equation}
   \chi^s = \frac{(\Gamma c)^2}{15}\bigg(\frac{\partial \beta_z(r)}{\partial r}\bigg)^2\Big(1+(\Gamma\beta_z)^2\Big)~,
\end{equation}
where $\Gamma$ and $ \beta_z $ are the bulk Lorentz factor and the dimensionless velocity component in the direction $z$ of the jet axis, respectively, and $r$ is the radial coordinate of the jet cross-section.
In order for the acceleration to be efficient, $ t_{acc}^s $ has to be shorter than the cooling time, here $ \tpg$, 
\begin{equation}\label{eq:shear_crit}
    \chi_{12}^sE_{p,12}\beta_\bot \gtrsim 6\times 10^{3}B_6~,
\end{equation}
where $ \beta_\bot $ is the dimensionless velocity component that is perpendicular to the magnetic field.
Since magnetized jets maintain $ B \gg 10^6 $ G, and since unmagnetized jets are prone to mixing which smears the shear gradient thereby reducing $ \chi^s$, the only possible shear acceleration site is the unmagnetized/weakly magnetized collimation shock. Our hydrodynamic simulations show that the shear layer between the shocked jet material and the cocoon can yield a typical velocity gradient from $ \Gamma \sim 100 $ to $ \beta \sim 0.1 $ over $ \Delta r \sim \theta_jR_s \sim 10^8 $ cm\footnote{Note that in our hydrodynamic simulation the jet is injected at $ \sim 10^7 $ cm, in principle the conical acceleration can start at $ r \approx r_g $, thereby having $ \Delta r $ that is smaller by roughly an order of magnitude, such that $ \chi_{12}^s \approx 10^2 $.}, which leads to $ \chi_{12}^s \approx 1 $. Consequently, we require $ E_p \gtrsim 10^4 $ TeV to satisfy Eq. \ref{eq:shear_crit}, however
at these energies and for $ B \approx 10^6 $ G, the Larmor radius becomes $ \gtrsim 10^7 $ cm, larger than the transversal confinement size. We conclude that neutrino production by shear acceleration is unlikely, as at low energies the shear acceleration timescale is longer than the cooling time, and at high energies the Larmor radius becomes too large.

\subsection{Neutrino spectrum}\label{sec:successful}

Following \S\ref{sec:magnetization} and \S\ref{sec:simulations}, collisionless subshocks may form if the radiation-mediated shocks are mildly magnetized, $ \sigma_u \gtrsim 0.1 $, as we find at the collimation shock in model $ \WM $ and internal shocks in model $ \HM $.
If the jet is launched hydrodynamically, a collisionless shock can only form after jet/cocoon breakout.
For highly magnetized jets, $ B' \gtrsim 10^{11} $ G at the collimation nozzle, which renders the neutrino production inefficient as the synchrotron cooling timescale is much shorter than their decay time, $ \fsupp \ll 1$.
We thus consider neutrino emission from the following sites: collimation shock in weakly magnetized jets (white ellipses in Fig. \ref{fig:successful_maps}k,m), internal shocks from highly magnetized jets (e.g. black circles in Fig. \ref{fig:successful_maps}a,d), and shock breakout from choked and successful hydrodynamic jets (Fig. \ref{fig:shock_breakout} for a successful jet breakout). In 
Appendix \ref{app:shock_breakout}, we describe the calibration of the latter by GRB170817 observations.

\begin{figure}
		\centering
		\includegraphics[scale=0.22]{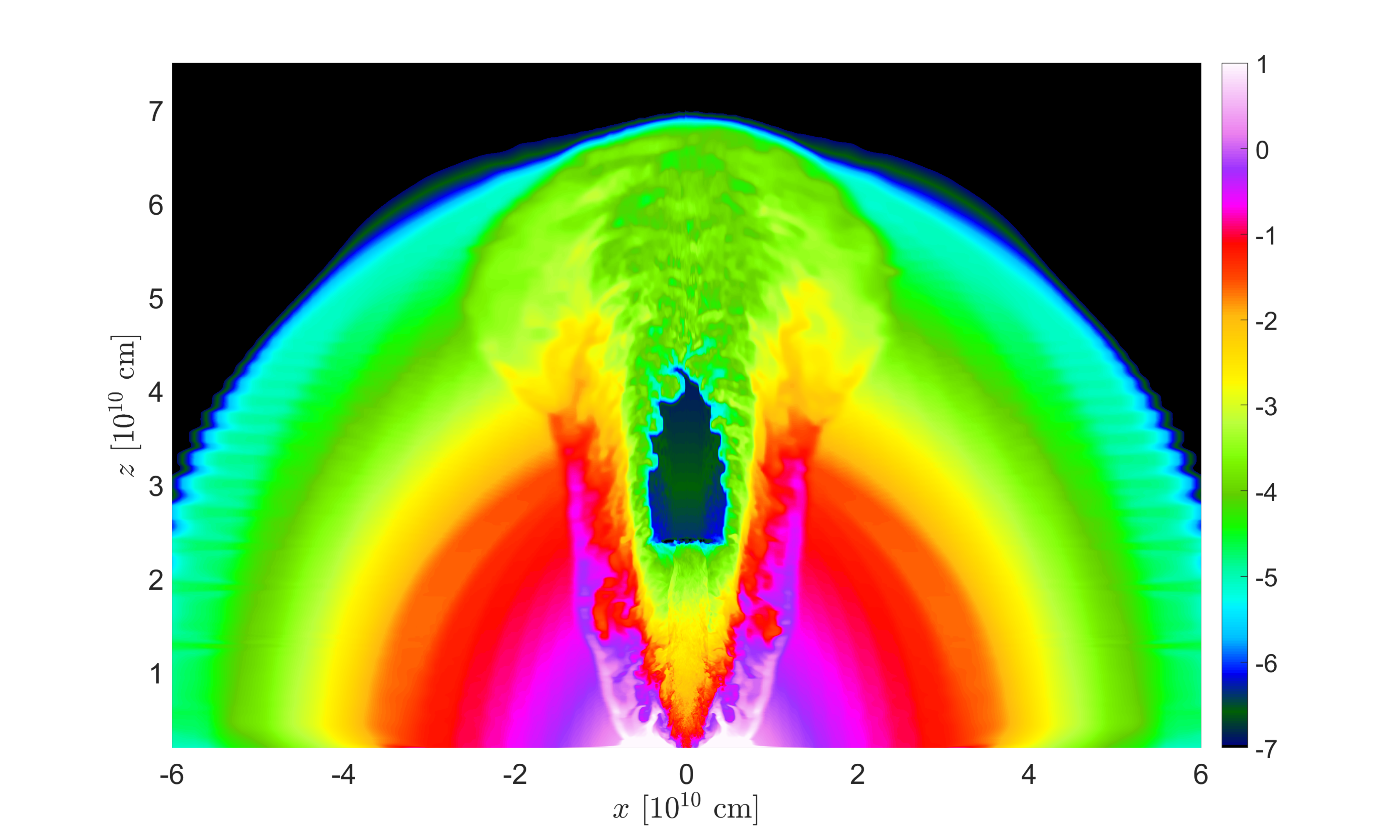}
		\caption[]{Logarithmic density map $ {\rm log}_{10}(\rho)~[{\rm g~cm^{-3}}] $ of the jet breaking out from the ejecta in the hydrodynamic model $ \SUa $, 2.8~s after the jet was launched. Since in this simulation the jet engine was working for $ t_j = 2 $ s, the unshocked jet material is at $ z = 0.8 {\rm s}\times c =2.4 \times 10^{10}~{\rm cm}$.
		}
		\label{fig:shock_breakout}
\end{figure}

\begin{figure*}
		\centering
		\includegraphics[scale=0.23]{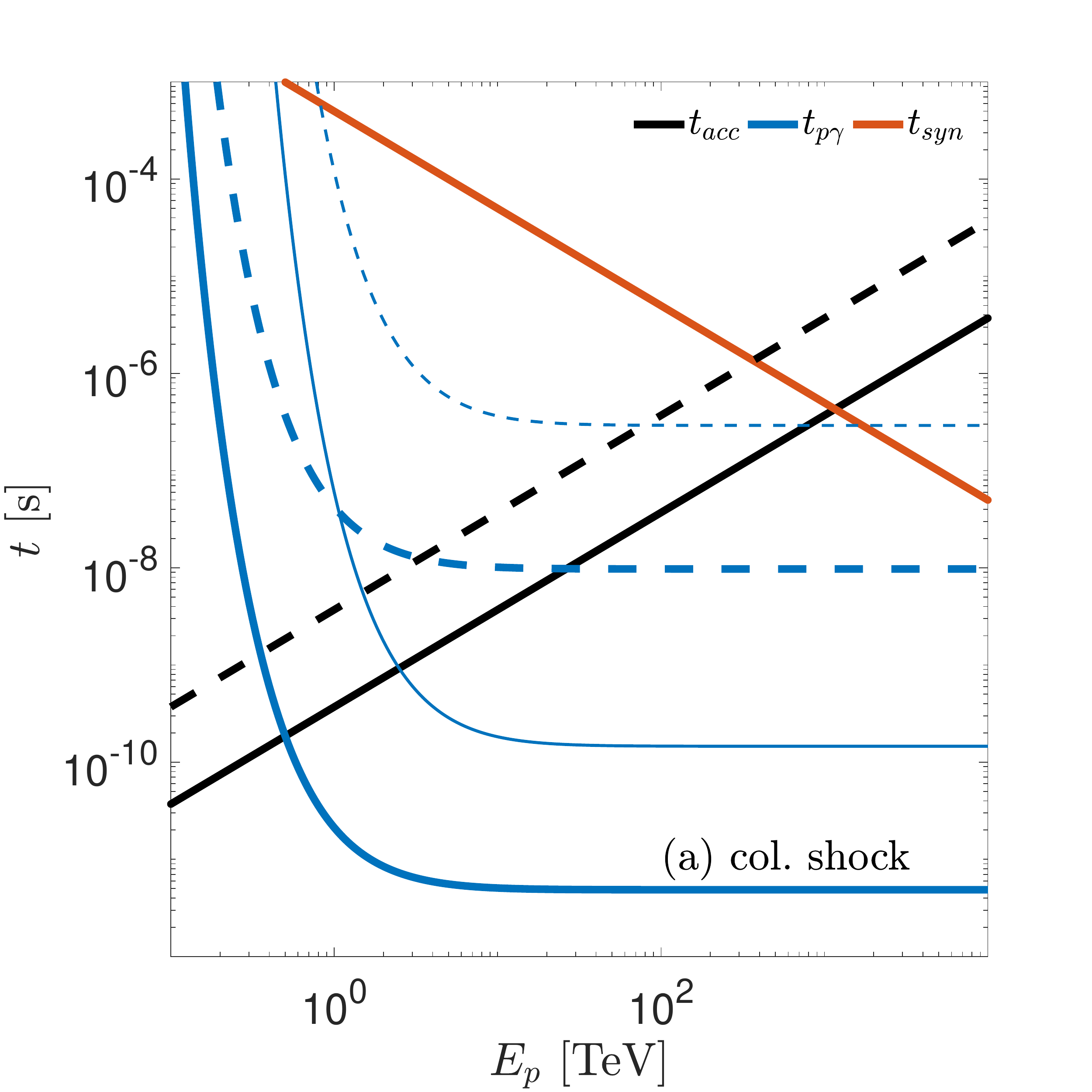}
		\includegraphics[scale=0.23]{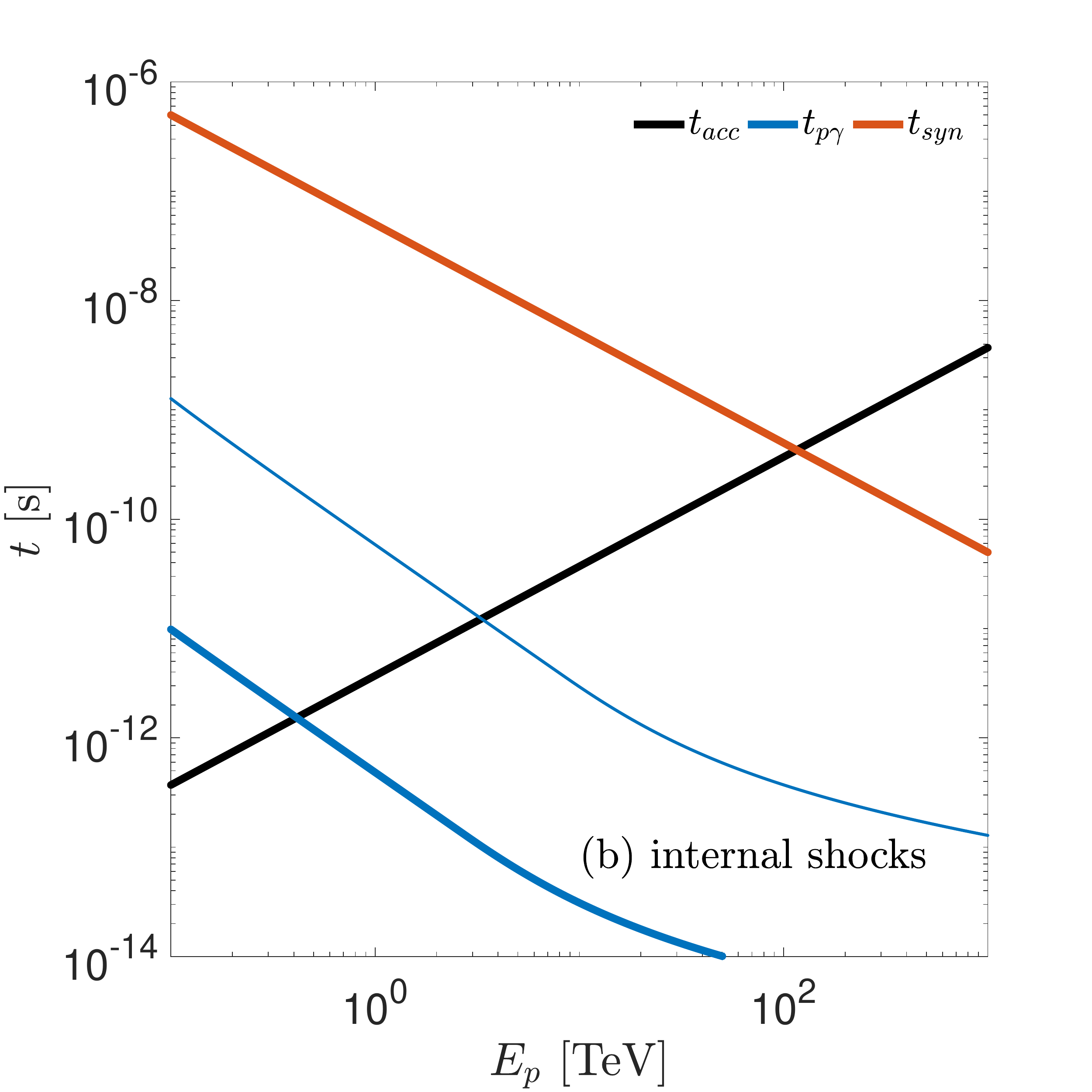}
		\includegraphics[scale=0.23]{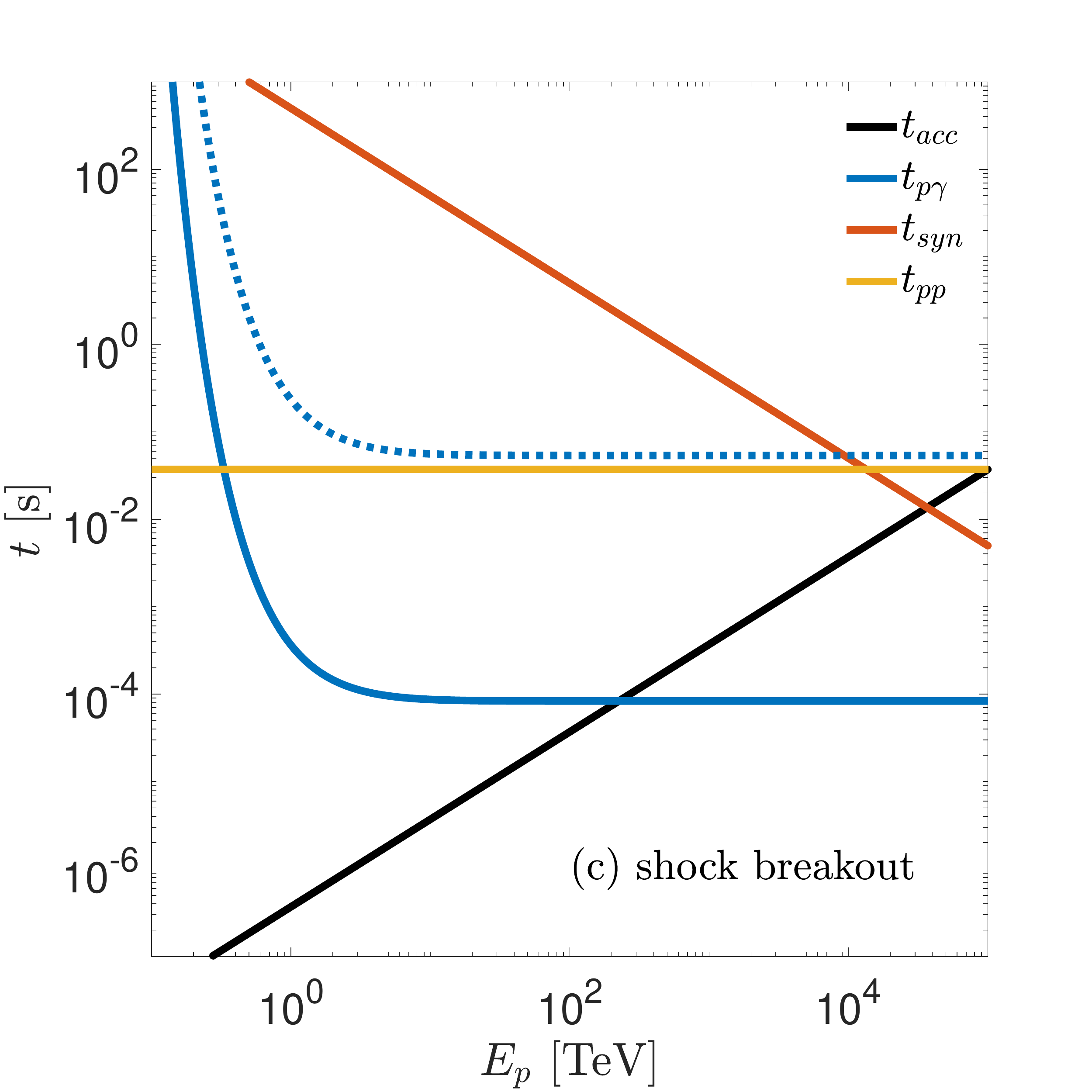}
		\caption[]{
		The acceleration (black) and cooling ($ p\gamma $ in blue and synchrotron in red) times in successful jets, at the following production sites for highly magnetized (HM) and weakly magnetized (WM) jets: collimation shock (WM) at different heights, $ 3\times 10^9 $ cm in solid lines and $ 3\times 10^{10} $ cm in dashed lines (panel a); internal shocks (HM, panel b); and shock breakout (hydrodynamic jets, panel c). The thin blue lines in panels a,b reflect the $ p\gamma $ cooling time when using the cross section for kaons, showing that $ p\gamma $ dominates for collimation and internal shocks also for kaons (the synchrotron time in panel a is for the shock at $ 3\times 10^9 $ cm, at $ 3\times 10^{10} $ it becomes negligibly long). In panel c we also show the $ pp $ cooling time (yellow) where we assume that the ejecta is composed only of protons. The typical $ p\gamma $ cooling time is shown both for the jet (solid blue) and for the cocoon (dotted blue).
		}
		\label{fig:nu_cooling}
\end{figure*}

\begin{figure*}
		\centering
		\includegraphics[scale=0.38]{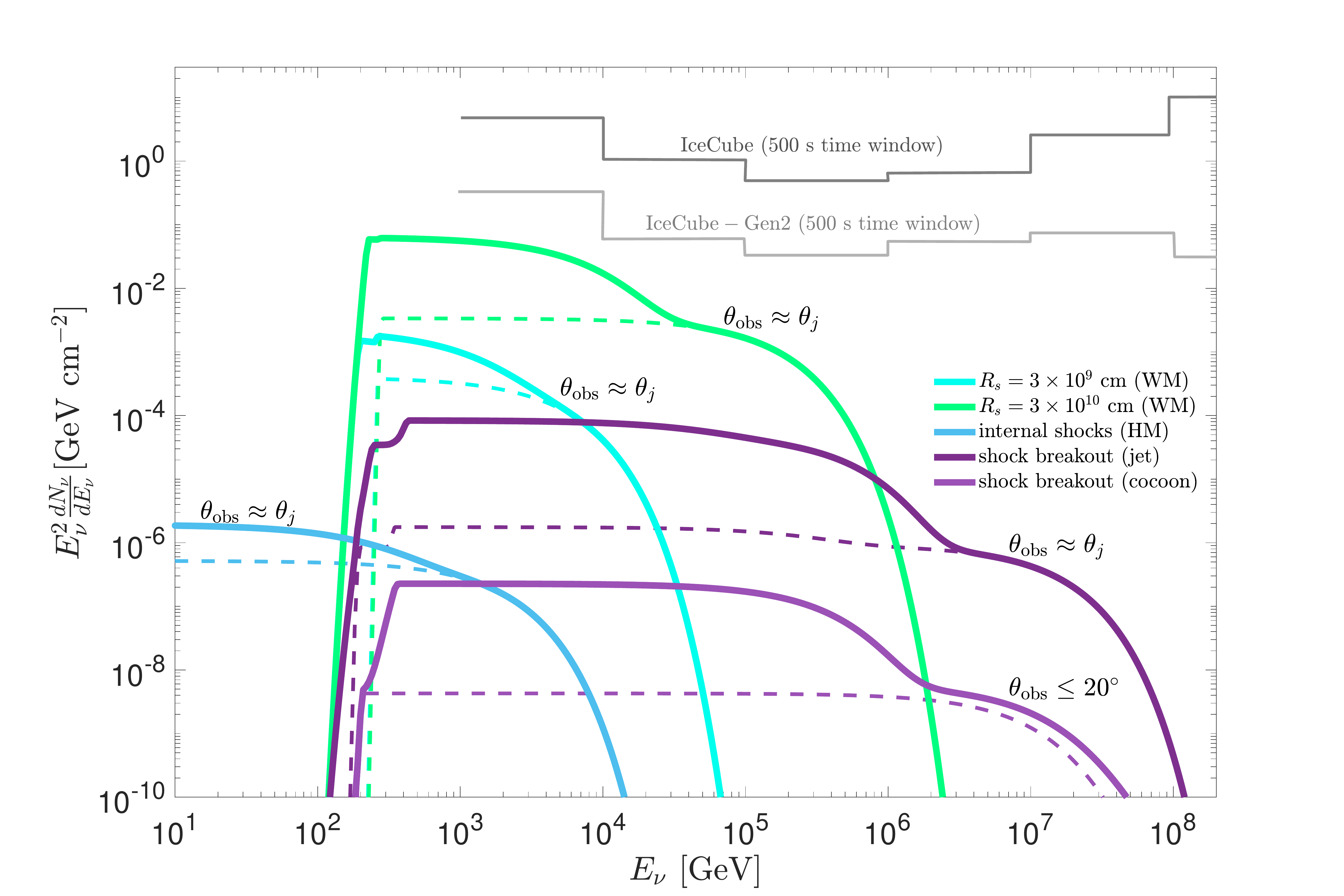}
		\caption[]{
		Neutrino signals from a BNS merger at 41 Mpc. Shown are the neutrino spectral fluences from the collimation shock at different heights (green and turquoise), internal shocks (light blue), which originate in the jet and can be observed within the jet opening angle. In purple is shown the shock breakout characteristic emission from the jet and the cocoon at viewing angle $ \theta_{\rm obs} = 20^\circ $ (successful and choked jet simulations produce similar neutrino emission from the cocoon).
		Solid lines mark the total neutrino emission and dashed lines mark the contribution from the decayed kaons.
		Upper limits (at 90\% CL) from various instruments on the neutrino spectral fluence from GW170817 during a $\pm 500$ s window centered on the gravitational wave (GW) trigger time are shown in grey and taken from \citet{aartsen2020}. All fluences are shown as the per-flavor sum of neutrino and anti-neutrino fluence, assuming equal fluence in all flavors, as expected for standard neutrino oscillation parameters.
	    At production sites where we use a Wien spectrum (collimation shock and shock breakout), the fluence rises at $ \sim 10^2 $ GeV due to the maximal photon energy, and falls at the maximal CR energy, as shown in panels (a-c). For internal shocks, where we use a Band function, the neutrino fluence is quasi-flat at low CR energies.
		}
		\label{fig:nu_spec}
\end{figure*}

Fig. \ref{fig:nu_cooling} depicts the acceleration (black) and relevant cooling times as a function of the proton energy in the comoving fluid frame at the following sites: collimation shocks for: $ R_s = 3\times 10^9 $ cm (solid) and $ R_s = 3\times 10^{10} $ cm (dashed) (panel a); internal shocks (panel b); jet (solid) and cocoon (dotted) shock breakout (panel c).
  The $ p\gamma $ cooling time (charged pions in thick lines and kaons in thin lines) is dominant at all production sites, both for pions and kaons, and thus the suppression factor is always of order unity. We note that at later times of the shock breakout, the photons density drops, causing the $ p\gamma$ process to become subdominant.
 
 	\begin{table*}
		\setlength{\tabcolsep}{3.5pt}
		\centering
		\begin{tabular}{ | l | c | c c |c c| c c| c c | }
			\hline
			 &  & \multicolumn{2}{c|}{Hydrodynamic} & \multicolumn{2}{c|}{weakly magnetized} & \multicolumn{2}{c|}{Highly magnetized} & \multicolumn{2}{c|}{Choked$^\dagger$} \\ \hline
			Acceleration & Production site & Peak fluence & Max. $ E_\nu $ &
			Peak fluence & Max. $ E_\nu $ &
			Peak fluence & Max. $ E_\nu $ &
			Peak fluence & Max. $ E_\nu $ \\
			& & $\left[{\rm \frac{GeV}{cm^2~s~sr}}\right]$ & [$ {\rm GeV} $] &
			$\left[{\rm \frac{GeV}{cm^2~s~sr}}\right]$ & [$ {\rm GeV} $] &
			$\left[{\rm \frac{GeV}{cm^2~s~sr}}\right]$ & [$ {\rm GeV} $] &
			$\left[{\rm \frac{GeV}{cm^2~s~sr}}\right]$ & [$ {\rm GeV} $]
			\\\hline
			\multirow{3}{*}{Fermi} & col. shock & \multicolumn{2}{c|}{-} & $ 10^{-1} $ & $ 10^3 $ & \multicolumn{2}{c|}{-} & \multicolumn{2}{c|}{-} \\
			& internal shocks & \multicolumn{2}{c|}{-} & $ 10^{-6}$&$10^3$&$ 10^{-6}$&$10^3$ & \multicolumn{2}{c|}{-}\\
			& shock breakout & $10^{-4}$&$10^5$ & $10^{-4}$&$10^5$&$10^{-4}$&$10^5$&$ 10^{-7}$&$10^5$ \\
			\hline
			Shear acceleration & \multicolumn{9}{c|}{-}  \\
			\hline
			
		\end{tabular}
		\hfill\break
		
		\caption{
		Prospects of neutrino maximal fluence at $ 41 $ Mpc, for different types of jets at different production sites. In sites where neutrino production is possible, we show the maximal neutrino energy and flux.
		\newline
		{$^\dagger$} Choked jets refer to the production of neutrino {\it after} the jet is choked.
		}
		\label{tab:neutrino}
	\end{table*}
  
  The resulting neutrino fluence from a GW170817-like event at GW170817 distance of $ D = 41 $ Mpc
 \citep{Hjorth2017} is shown in Fig.~\ref{fig:nu_spec}. A summary of the neutrino peak fluence for different types of jets and at different production sites is given in Table~\ref{tab:neutrino}.
 It is shown that the neutrino signals are composed of two components: a strong low energy component and a weaker high energy tail, which correspond to the pion and kaon contributions, respectively. The latter emerges thanks to the smaller $ p\gamma $ cross section of kaons which allows CR to accelerate to higher energies prior to the $ p\gamma $ interaction (Fig. \ref{fig:nu_cooling}a,b). Since kaons are heavier than pions, they have a longer synchrotron cooling time by $ \sim 2 $ orders of magnitude. When the magnetization is high, the pions cool down rapidly, and thus the kaon contribution becomes important at all neutrino energies.
 
 Upon launching, the high magnetization and photon density prevent the production of high energy neutrinos. At later times, the magnetic field at the internal shocks drops, but so is their magnetization such that their ability to support a strong collisionless subshock is questionable.
 If the jet engine is working for a long time ($\gtrsim 2$~s) and the jet initial magnetization is mild, then a large ($ \sim 10^{10} $ cm) collimation shock forms, where pions decay before undergoing synchrotron cooling. This results in our highest fluence of neutrinos up to $ \sim 10 $ TeV. However, such a scenario is unlikely as it requires jet launching with subdominant magnetic fields, and atypically long duration of jet launching.
 
 The neutrinos emitted from the shocks inside the ejecta are  confined to the jet opening angles.
 In the shock breakout scenario, although the strongest neutrino signal emerges from shocks inside the jet, a weaker signal also emerges from the shock breakout of the mildly relativistic cocoon, at viewing angles $ \lesssim 20^\circ $. The latter is also a robust estimate for the neutrino fluence, as it is calibrated by GRB170817A observations, and weakly dependent on the jet magnetization and fate (successful and choked jets have similar neutrino emission from the cocoon). Shock breakout is also the only neutrino production site for hydrodynamic jets, or choked jets (see Appendix~\ref{app:choked}).

\section{Cosmic-ray production}\label{sec:CR}

We consider the CR acceleration at the external shock wave. Once the structured jet (or cocoon, if the jet is choked) completes the initial acceleration phase, it is coasting freely in the interstellar medium (ISM). Assuming the ISM has a constant number density $n_0$, and assuming an independent hydrodynamic evolution of each angle $ \theta $, the deceleration spherical radius of the shock is
\begin{equation}\label{eq:r_dec}
    r_{dec}(\theta) = \bigg(\frac{3M(\theta)}{4\pi n_0 m_p\Gamma(\theta)}\bigg)^{1/3}~,
\end{equation}
where $ M(\theta) $ is the isotropic equivalent swept mass of the outflow at angle $ \theta $.
As the shock decelerates, it deposits a fraction of its dissipated energy into accelerating electrons in the magnetic field generated by the shock. The accelerated electrons emit synchrotron radiation to form the afterglow signal. In the following, we use a numerical model that fits GW170817A's afterglow observations, as found by \citet{Mooley2018b}. We also consider a case of an uncollimated choked jet (models $ \SU $ and $ \CU $ in Appendix~\ref{app:simulations}). Fig.~\ref{fig:homologus} depicts maps of kinetic energy density (left panels) and $\Gamma\beta$ (right panels) for these models.

The maximal energy at which CR can be accelerated at the shock is estimated by equating the acceleration time $t'_{acc}= t'_L/\xi_{acc}= E'_p/(ZqB'c\xi_{acc})$ with the escape time $ t'_{esc} = R'/(\Gamma\beta c) $ \citep{Hillas1984}, where $R' $ is the proper acceleration length, $ q $ is the elementary charge, and $ \xi_{acc} \approx 0.1 $ is the efficiency of the acceleration in the mildly relativistic case \citep[e.g.,][]{globus2015grb}. We find
\begin{eqnarray}
\Epmaxprime(\theta)&=&\frac{\xi_{acc}R'(\theta) Z q B'(\theta)}{\Gamma(\theta)\beta(\theta)}\nonumber\\ &\approx&300\,\xi_{acc}Z B'_{\rm G}(\theta) R_{\rm cm}(\theta)\left[\Gamma(\theta)\beta(\theta)\right]^{-1}\, {\rm eV}\,.
\label{emax_conf}
\end{eqnarray}
Note the angular-dependent quantities since we perform a semi-analytic calculation on structured jets. We assume the acceleration length to be the proper shock width, $R' \approx {r_{dec}}/{12\Gamma} $ \citep[e.g.,][]{vanEerten2015}, and use the general convention that the magnetic energy density behind the shock can be characterized by the equipartition parameter $ \epsilon_B $, such that the magnetic field is $ B'\approx \left[32\pi\epsilon_Bn_0m_p\Gamma(\Gamma-1)\right]^{1/2}c $. Plugging $ B' $ and Eq.~\ref{eq:r_dec} into Eq.~\ref{emax_conf}, the maximal energy of the accelerated particle is:
\begin{equation}\label{eq:max_CR}
\footnotesize
    \Epmaxprime(\theta) \approx 2\times 10^7~\frac{\xi_{acc}Z(\Gamma(\theta)-1)^{1/2}}{\beta(\theta)}\left(\frac{n_0 M_{-6}(\theta)^2\epsilon_{B,-3}^3}{\Gamma(\theta)^{11}}\right)^{1/6}~{\rm GeV}~\,,
\end{equation}
where as before, $ Q_x $ denotes the value of the quantity $ Q $ in units of $ 10^x $ times its c.g.s. units, except for $ M_x $ which is given in units of solar masses.

As the physical quantities have an angular dependence (see Fig.~\ref{fig:homologus}), we discretize the shock wave into bins in $\theta$ and $ \Gamma $ (on the radial dimension), on which the quantities do not vary much.
We assume that the angles can be treated independently of each other (e.g., no spreading), and that the mass in the front decelerates to the Lorentz factor of the mass that supports it behind. That is, the cumulative mass from radius $ r $ is moving with $ \Gamma(r) $.
We calculate the cumulative isotropic equivalent mass over the radial direction to calculate the maximal CR energy from Eq.~\ref{eq:max_CR} for each angle and proper-velocity, which corresponds to different times. We then use Eq.~\ref{eq:CR_spec}\footnote{Note that here we perform a full calculation at each angle, and thus we use the total energy rather than the isotropic equivalent energy.} to calculate the comoving CR spectrum at each angle and proper-velocity.
We boost the rest-frame spectra to the observer by accounting for the beaming and the Doppler boost for each observer. We ignore the temporal evolution of the emission, and integrate over time (radial direction) since the CR diffusion time is longer than the emission times. Finally, we average the angle-dependent flux over the beaming factor as CR do not propagate in straight lines in the Galactic magnetic field, and obtain the CR spectrum from a single source.

\begin{figure}
		\centering
		\includegraphics[scale=0.2]{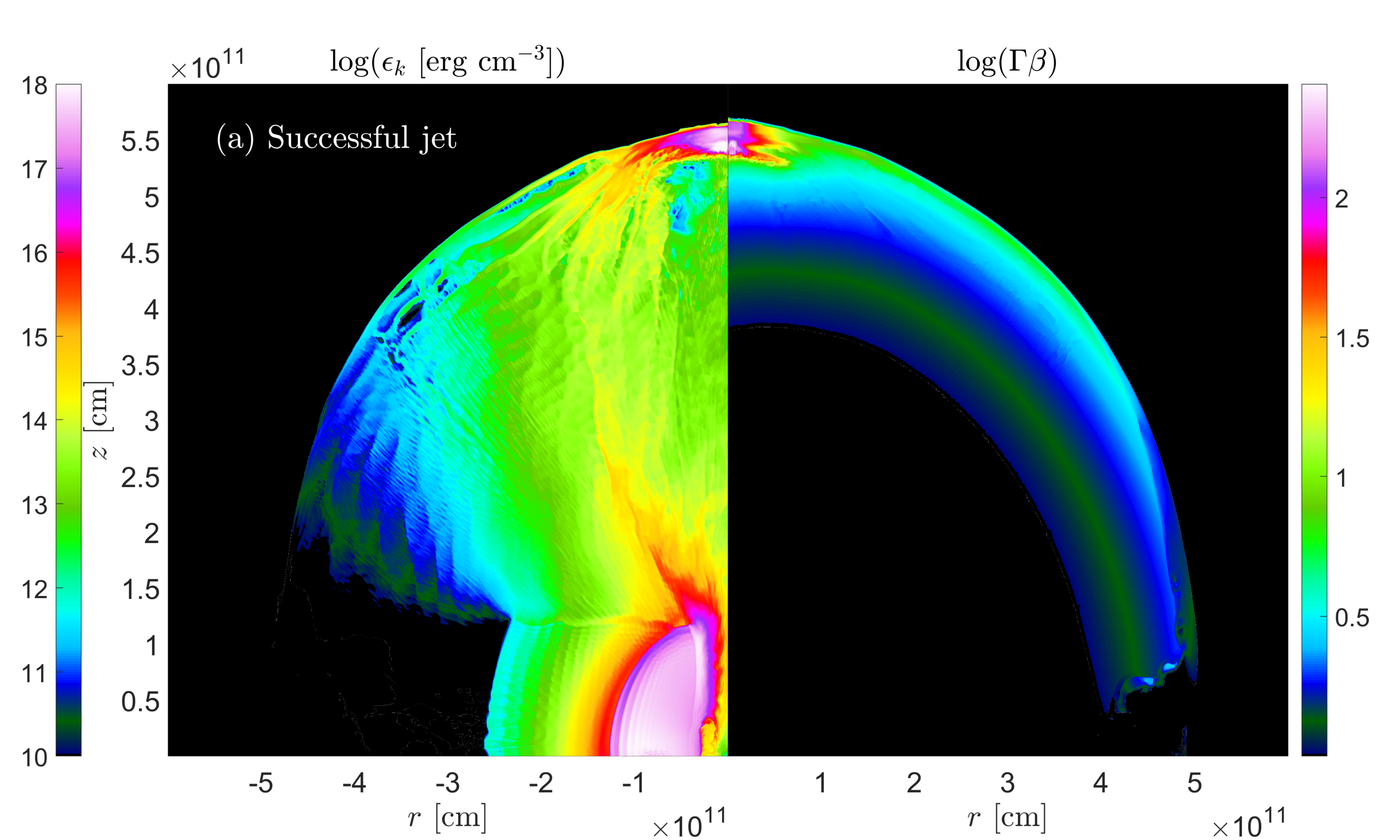}
		\includegraphics[scale=0.2]{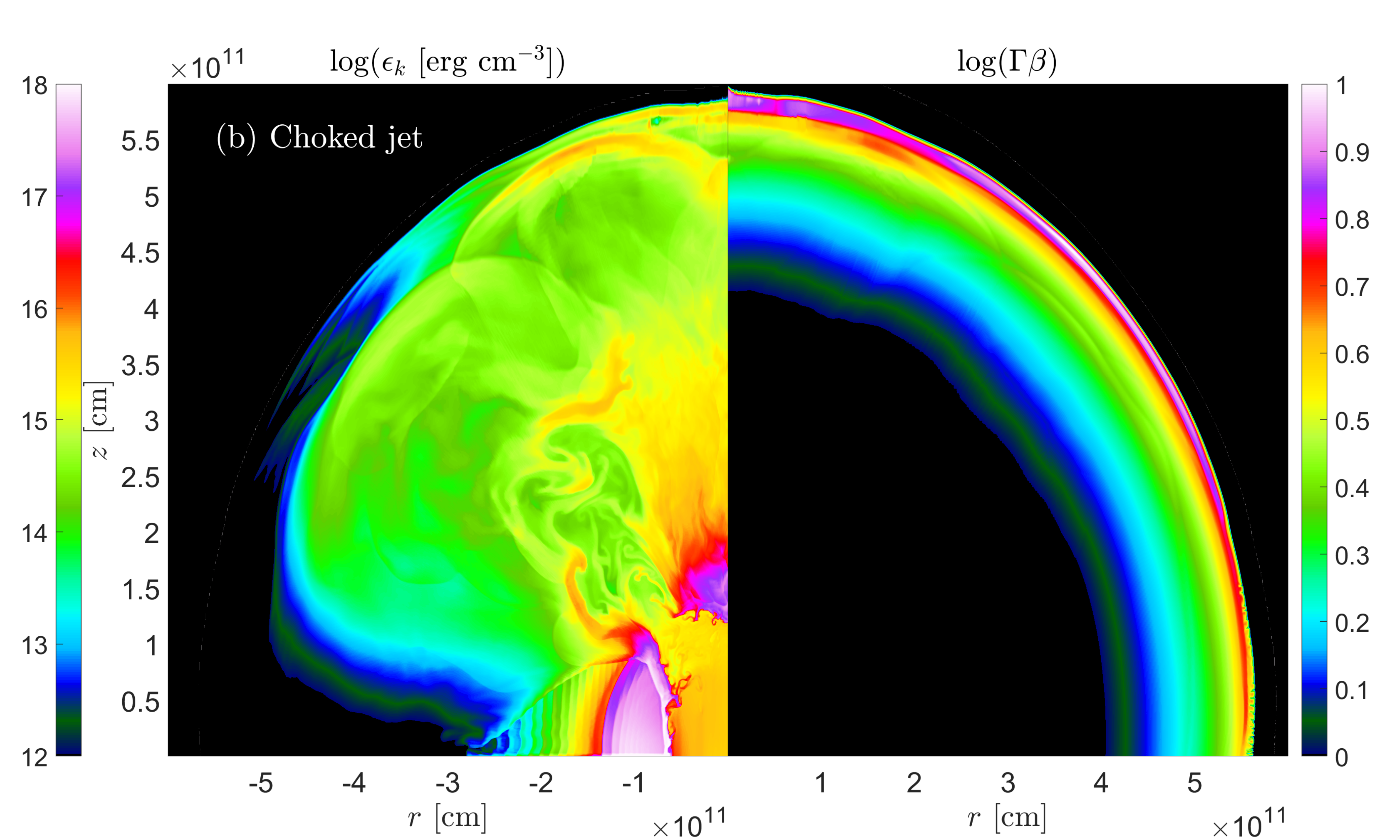}
		\caption[]{
            Maps of successful (a) and uncollimated choked (b) jets during the free coasting phase. Shown are the logarithmic kinetic energy density (left) and the logarithmic proper-velocity (right).
		}
		\label{fig:homologus}
\end{figure}

   \begin{figure}
   		\centering	\includegraphics[scale=0.38]{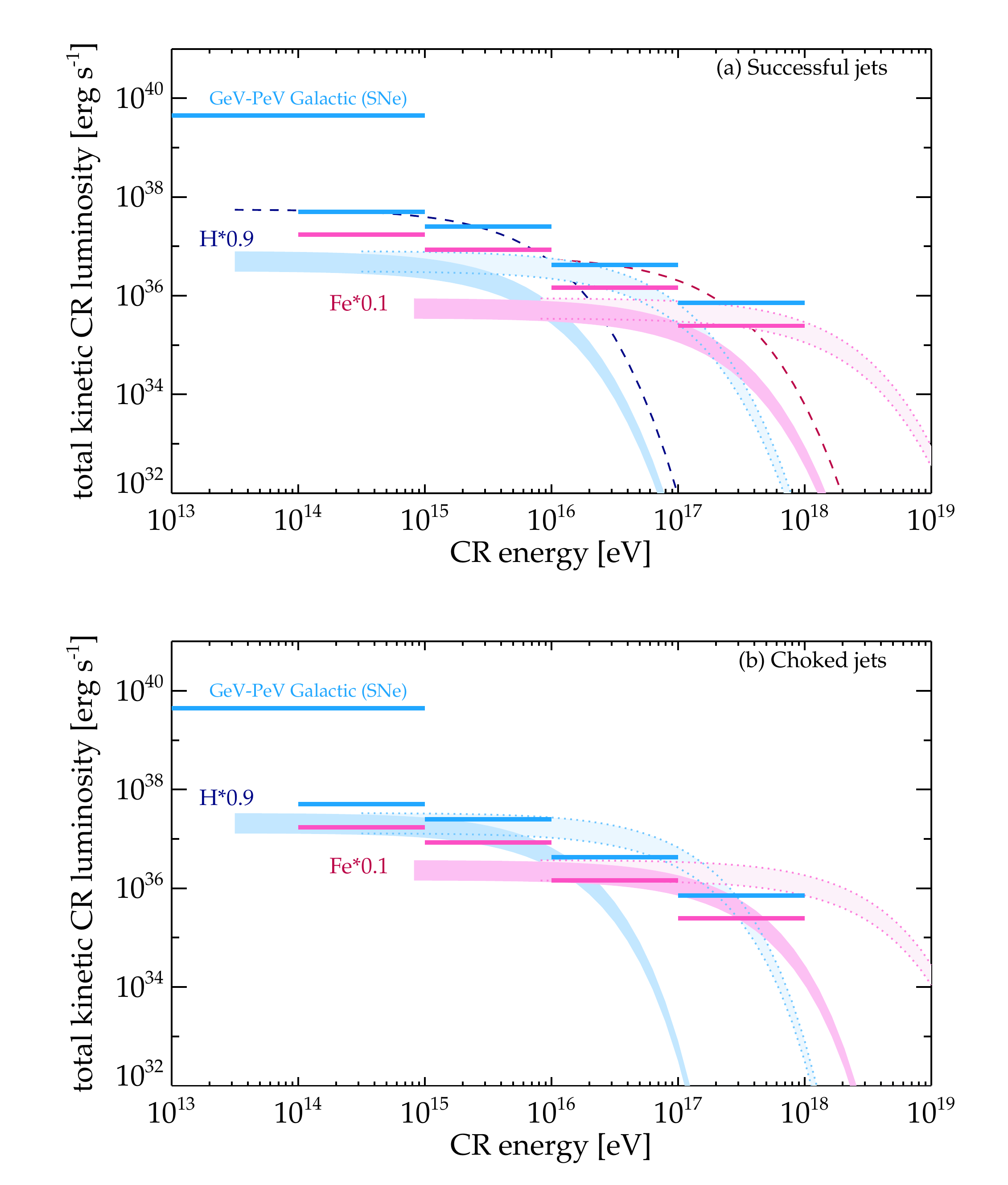}
    		\caption[]{Total CR luminosity in successful (a) and choked jets (b) from external shocks from BNS mergers, assuming that 90\% of the energy dissipated into CR is deposited in light components (protons, in blue), and 10\% is in heavy (Fe, in pink) elements. We assume that the rate of both successful and choked jets is the GBNS merger rate ${\cal R}_{MW}=42^{+30}_{-14}$ Myr$^{-1}$. We use $\xi_{acc}=0.1$, $\epsilon_{CR}=0.1$, and two values of $ \epsilon_B $, that of GW170817, $\epsilon_{B,-3}=1$ (strong colors), and the more commonly used $\epsilon_{B,-3}=100$ (faint colors). The horizontal bars denote the Galactic luminosity in CR in different energy bins, estimated for protons and iron. In the successful jet scenario (calibrated by GW170817), the BNS account for 5-10\% of the total Galactic CR luminosity. This also leads to an upper limit of the GBNS merger rate, ${\cal R}_{MW}=5\times 10^{-4}$ yr$^{-1}$ (dashed lines).
    		For the choked jets, the CR luminosity is higher, but poorly constrained due to the unknown rate and energetics of such events.
    		}
    		\label{fig:CR_spec}
    \end{figure}

To estimate the total CR luminosity available from GBNS mergers, we use the rate constrained by the  known sample of GBNS systems, ${\cal R}_{MW}$ ($42^{+30}_{-14}$ Myr$^{-1}$; \citealt{pol2019}), in agreement  with LIGO estimates $ \RBNS = 320^{+490}_{-240}~{\rm Gpc}^{-3} {\rm yr}^{-1}$ \citep{abbott2021}, which we assume to be the rate of both successful jets and choked jets.
Using  $\epsilon_{CR}=0.1$, $\xi_{acc}=0.1$, and  circum-merger number density $n_0=1\, {\rm cm}^{-3}$ \citep{Berger2014}, we calculate the CR luminosity from BNS mergers, assuming that protons constitute 90\% of the energy and Fe account for the remaining 10\% \citep[see][for a discussion about the CR composition from the kilonova ejecta]{Kimura2018b}. Note that we do not aim to accurately fit the composition as this depends on many parameters (e.g., the jet parameters, microphysics of the shock and BNS merger rate), but rather compare the CR energy from our model with the observed CR in the Milky Way, which is estimated as follows. 
The kinetic energy density of cosmic rays in the Milky Way is $\rho={4\pi}{c}^{-1}\int \Phi(E)dE$.
Up to the knee, the Galactic CR flux is $\Phi(E)\sim1.8 \times 10^4(E/1 { \rm GeV})^{-1.7}\rm m^{-2} s^{-1} sr^{-1}$ and $\rho_{\rm GeV-PeV}\sim1 \rm eV\, cm^{-3}$. Above the knee,  $\Phi(E)\sim1.8\times 10^{-6}(E/1 {\rm PeV})^{-2.1} \rm m^{-2} s^{-1} sr^{-1}$ and $\rho_{\rm PeV-EeV}\sim 6.8\,10^{-5} \rm eV\, cm^{-3}$. The measured ratio of radioactive isotopes and the secondary-to-primary ratios indicate a residence time of CR in the Galactic disk of volume $V\sim{\pi(15 \rm kpc)^2}(200 \rm pc)$,  of the order $\tau\sim 30\, {\rm Myr}\left[{(E/Z)}/{\rm 5 GeV}\right]^{-\delta}$ with $\delta\sim 0.33$. Assuming protons, $Z=1$, the power required to maintain the observed Galactic cosmic ray flux in the GeV-PeV range is $\rho V/\tau\sim 4.5\times 10^{39}~{\rm erg\,s}^{-1}$, a few percent of the total kinetic power of supernovae. Above the knee, in the PeV-EeV energy range, the required power is only $\sim 2.7\times 10^{37}~{\rm erg\,s}^{-1}$ but the sources need to accelerate CR to higher energies than typical supernovae remnants ($\sim100$ TeV).

Fig.~\ref{fig:CR_spec} depicts the resulting CR luminosity for BNS merger  rate ${\cal R}_{MW}=42^{+30}_{-14}$ Myr$^{-1}$ and $\epsilon_{B,-3}=1$ (dark shaded area) as constrained by GW170817's afterglow \citep{Mooley2018b}, and $\epsilon_{B,-3}=100$ (light shaded area). We find that blast-waves of sGRBs may account for $\sim5-10$\% of the CR luminosity in the Milky Way in the PeV-EeV energy range for GW170817-like events if the jet is present.
The maximum energy for the iron component, $\sim2\times 10^{17}$ eV ($10^{18}$ eV) for $\epsilon_{B,-3}=1$ ($\epsilon_{B,-3}=100$).
Choked jets seem to be able to account for all Galactic CR flux, owing to their quasi-spherical nature which increases the angle averaged emission\footnote{We also note that in our simulation the uncollimated choked jet has 2.5 times more energy than that of GW170817 (Appendix \ref{app:simulations}).}. However, we stress that this result is poorly constrained as we assume the rate and the energy of choked jets to be comparable to that of successful jets, but those quite possibly are lower for choked jets.

The dipole anisotropy for a single transient source has an amplitude $\Delta=3r/2ct$ for an event  occurring at a time $t$ in the past and located at a distance $r$. The upper limit on the observed dipole anisotropy in the PeV-EeV range is $\Delta\sim0.02$.
If the latest GBNS occurred at $1/{\cal R}_{MW}$ in the past, with ${\cal R}_{MW}=42^{+30}_{-14}$ Myr$^{-1}$, its distance would be in the range 0.6-1.5 kpc (1-3 kpc) if the BNS mergers account for 10\% (5\%) of the Galactic CR flux in the PeV-EeV energy range. 

\section{Summary of results}\label{sec:summary}

We performed one GRMHD simulation of a highly magnetized jet and additional RMHD simulations of a weakly magnetized jets to estimate neutrino and CR signals from hydrodynamic/magnetized and choked/successful jets.
Our main findings are:\\
    $\bullet$ collisionless shocks cannot form below the photosphere due to mixing between jet and cocoon material; collisionless subshocks can form under certain conditions that depend on the jet magnetization;\\
     $\bullet$  hydrodynamic jets and choked jets cannot produce neutrinos prior to jet-breakout;\\
     $\bullet$ if the jet is operating for atypically long time for sGRBs ($ \gtrsim 2 $~s) and its magnetization is mild at the collimation shock, the collimation shock can produce $ \sim 10 $ TeV neutrinos; \\
     $\bullet$  internal shocks, for which we use those that emerge in the full GRMHD simulation, yield a signal well below detection limits;\\
     $\bullet$  jet/cocoon shock breakout, which is considered here as a source of neutrino in BNS for the first time, is independent of the jet magnetization and the only production site for hydrodynamic jets. It can produce $ \sim 10 $ PeV neutrinos to large angles, on order of the cocoon opening angle of $\sim20$\degree; however, the emission is $\sim100$ times lower than collimation shocks at 1~PeV and would require events within our local group or more sensitive neutrino detectors;\\
     $\bullet$  the shear layer does not produce high energy neutrinos;\\
     $\bullet$  external shocks, responsible for the afterglow emission, can accelerate CR up to $\sim10^{18}$ eV. 
     
     To summarize, we find that sGRBs do not produce a neutrino signal that is detectable by current-day facilities. We find that GBNS can account for 5-10\% of the Galactic CR flux in the PeV-EeV range, depending on the GBNS merger rate, jet energetics and CR composition.

\section{Discussion}\label{sec:discussion}

 As long as the jet is propagating inside the ejecta, its mixing with the cocoon plays a crucial role in the jet dynamics, and inhibits the formation of collisionless subphotospheric shocks in the jet. If the jet maintains some non-negligible degree of magnetization, collisionless subshocks may form, allowing CR to be accelerated efficiently if the Larmor radius is larger than the subshock width.
 We found that the strongest neutrino flux emerges from the collimation shock in the weakly-magnetized jet, however its existence is highly questionable due to the atypical engine work-time and the  fine-tuned magnetization.
 Internal shocks suffer from heavy mixing at early times, when the jet is still sufficiently magnetized to support the formation of collisionless shocks, and thus do not yield a promising neutrino signal either.

Our results are in agreement with those of \citet{Biehl2018} who found that neutrino signals from sGRBs are unlikely to be detected by current-day facilities, and with \citet{Kimura2018a} who found that collimation shocks of sGRBs are unlikely to produce a detectable signal.
However, these studies, which have addressed the neutrino emission analytically and thus overlooked the effects of mixing and jet magnetization on neutrino production at subphotospheric shocks, also found internal shocks to be the most promising neutrino production sites.
We showed that in realistic jet-cocoon outflows, the production of neutrinos at subphotospheric shocks may only be possible if the jet is magnetized. Our magnetized jet simulations show that the required magnetization is only present at small radii, where the high photon density prevents CR from efficiently accelerating to high energies. Consequently, we find that it is unlikely that subphotospheric internal shocks provide any detectable neutrino signal.

The jet/cocoon shock breakout is considered here as a source of neutrinos in BNS for the first time. It is generally independent of the jet initial magnetization, and its energy is calibrated by GRB170817A observations. The neutrinos produced by the decay of kaons in the shock breakout can reach $ \sim 10 $ PeV, owing to low magnetization and photon density which allow protons to accelerate to high energies.
If the shock breakout is accompanied by a collimated jet, such that the outflow breaks out with an angular structure, the on-axis neutrino emission ($ \theta \lesssim \theta_j $) is $ \sim 2-3 $ order of magnitude brighter than that of the cocoon ($ \lesssim 20^\circ $), which arises regardless of the jet's fate.
The energy stored in the shock breakout is smaller than in the other potential production sites, and thus such emission may only become detectable in the future.
We find that the neutrino fluence from on-axis shock breakout is comparable with that from fallback accretion of unbound ejecta in BNS mergers \citep{Decoene2020}. Note however that neutrinos from fallback accretion are generated over much longer timescales, such that the neutrino flux from the jet breakout emission is much higher, and the flux from fallback accretion is comparable with that from the cocoon shock breakout.
Finally, unlike subphotospheric production sites where neutrinos are anticipated to arrive before the photons, neutrinos from shock breakout are candidates for a neutrino-photon coincident detection. At distances of $ \sim 100 $ Mpc, the arrival times of TeV-PeV neutrinos and photons to Earth is expected to be essentially the same \citep[e.g.,][]{Wei2016},
allowing a coincident detection of the two, which can be a probe of the neutrino production site.

Our estimate for the CR emission from the jet-cocoon blast wave, calibrated by GRB170817A observations, have several important implications. First, the Galactic CR luminosity sets an upper limit to the GBNS merger rate to ${\cal R}_{MW}\sim 5\times 10^{-4}$ yr$^{-1}$ for successful jets. Second, the level of anisotropy in the PeV-EeV range constrains the distance from the last GBNS merger. In a purely turbulent magnetic field of $\sim1 \mu G$, the typical angular spread of CR sources at distances $\lesssim$ 1 kpc is $\lesssim50$ degrees at 50 PeV, such that it might be possible to observe an intermediate-scale anisotropy from CR sources emitting in the shin region from BNS at such distances.
Even if  unlikely that the last BNS occured at such close distance, it is worth mentioning that at 33 PeV, an intermediate-scale anisotropy has been reported in the direction ($l\sim$80\degree, $b\sim$15\degree) in Galactic coordinates \citep[see figure 3 of][]{ahlers2019}, and owing to their natal kicks, we expect an anisotropy from a BNS merger to be located at higher latitudes than other Galactic transient sources.
In the more likely scenario that the latest BNS merger occurred at a larger distance,  our estimates suggest a distance of $\lesssim3$ kpc for the latest GBNS merger. Future combined search for CR anisotropy and X-ray line emission from radioactive elements \citep[e.g.,][]{wu2019} could pinpoint the latest BNS merger in our Galaxy.

Finally, our results also have a few interesting implications to long GRBs, which share similar jet-cocoon mixing and jet structure with sGRBs \citep{Gottlieb2021b}. Since long GRBs operate over longer engine work-times and thus are also more energetic, it is possible that such systems can produce more neutrinos, particularly in the larger collimation shock. Furthermore, these conditions may help achieving optically thin shocks as suggested by \citet{Murase2013}, with the possible extended envelope is a promising production site \citep{Nakar2015}.
It is also worth mentioning that the dynamics of long GRB jets propagating in their progenitor stars is not significantly different in terms of the emerging jet-cocoon structure \citep{Gottlieb2021a}. Since the energetics times the long GRB rate is higher than that of short GRBs, it implies that the contribution of long GRBs to the Galactic PeV-EeV CR flux is even higher than that of short GRBs.

\section*{Acknowledgements}
We are grateful to Amir Levinson, Anatoli Fedynitch, Kohta Murase, Andrew MacFadyen, Roger Blandford, Denis Allard, Brian Metzger, Alexander Tchekhovskoy and Matthew Liska for helpful discussions and useful comments.
OG acknowledges support by the Israel Science Foundation Grant No. 1114/ 17 and ERC grant (JetNS). NG acknowledges support by the Czech Science Foundation under the grant GACR 20-19854S titled “Particle Acceleration Studies in Astrophysical Jets”, partial support from the National Science Foundation through AST grant 1715661, and the Simons Foundation.

\bibliography{refs}

\appendix
\section{Numerical simulations setup}\label{app:simulations}

All simulations make use of a relativistic ideal gas equation of state, piecewise parabolic reconstruction method and an HLL Riemann solver.
In the RMHD simulations, the jet injection height is $ z_{\rm{beg}} $, and thus its injection nozzle radius is $ r_{\rm noz} = z_{\rm{beg}}\theta_j $, where the origin is the center of the system of the merger. The weakly magnetized jets are injected with a toroidal magnetic field whose peak magnitude at $ r = r_{\rm noz}/2 $ is set by $\sigma_0 = 10^{-2} h_j $, where $ h_j $ is the jet's initial specific enthalpy \citep[see][for the full profile]{Gottlieb2020b}. The jet quantities of each model are listed in Table \ref{tab:models}.
We employ 3D Cartesian grids to properly account for the formation of instabilities and avoid numerical artifacts \citep[see e.g.,][]{Gottlieb2021a}. After the jet reaches the homologous expansion phase, the hydrodynamic jets are remapped into axisymmetric grid for modeling the afterglow phase \citep[see][for details]{Nakar2018}. The grid setups of the successful hydrodynamical and weakly magnetized models are provided in \citet{Mooley2018b} and \citet{Gottlieb2020b} respectively\footnote{The hydrodynamic choked jets has an identical grid setup to the successful jets, and the weakly magnetized choked jet has an identical setup to the successful weakly magnetized jet with two differences: the jet launching duration $ t_j $, and a higher grid resolution by 25\% on $\hat{x} $- and $ \hat{y} $- axes and by 60\% in $ \hat{z} $-axis.}.

In the GRMHD simulation the vector-potential profile is
\begin{equation}
    A=A_\phi(r,\theta) = 10^{13}\frac{{\rm sin}\theta}{r}\cdot {\rm max}\bigg(\frac{r^2}{r^3 + r_c^3} - \frac{r_o^2}{r_o^3 + r_c^3},0\bigg)~{\rm Mx~cm^{-1}}~,
\end{equation}
where $ r_c = 5\times 10^7 $ cm and $ r_o = 10^{10}~{\rm cm} $. The second term sets $ B = 0 $ at $ r \approx 15 r_g $ and at $ r = r_o $.
The jet is generated self-consistently, its resulting parameters are listed in Table \ref{tab:models}.
We use a tilted metric to avoid numerical dissipation on the jet axis at $ \theta = 0 $. 
The GRMHD simulation is carried out in spherical coordinates with logarithmic grid on $ \hat{r} $-axis from $ r_g $ to $ 5\times 10^9 \cm $, and uniform on $ \hat{\theta} $- and $\hat{\phi}-$axes .
We use two levels of AMR, which is activated at high entropy to properly resolve the jet and the cocoon. Our effective resolution in the maximal refinement level is $ 1152 \times 576 \times 512 $ cells on $\hat{r}-\hat{\theta}-\hat{\phi} $ respectively, enough to resolve the development of magneto-rotational instabilities in the disk.

	\begin{table}[!h]
		\setlength{\tabcolsep}{11pt}
		\centering
		\begin{tabular}{| l | c c | c c c c c c c | }
			
			\hline
			 Model & $ \sigma_0 $ & Jet fate & $ t_d $ [s] & $ \theta_j $ & $ h_j $ & $ z_{\rm{beg}} $ & $ L_j [10^{50}~{\rm erg~s^{-1}}] $ & $ t_j $ [s] & $ \Gamma_\infty $ \\ \hline
			
			$ \SU $ & 0 & Successful & 0.2 & $ 4^\circ $ & 200 & $ 4.5\times 10^8~{\rm cm} $ & 1.4 & 0.8 & 2000 \\
			$ \SUa $ & 0 & Successful & 0.4 & $ 15^\circ $ & 100 & $ 3\times 10^8~{\rm cm} $ & 2 & 2.0 & 300 \\
			$ \SUb $ & 0 & Successful & 0.6 & $ 8^\circ $ & 100 & $ 7\times 10^8~{\rm cm} $ & 0.3 & 3.6 & 500 \\
			$ \CU $ & 0 & Choked & 0.2 & $ 23^\circ $ & 20 & $ 4.5\times 10^8~{\rm cm} $ & 3 & 1.0 & 32 \\
			$ \WM $ & $ 10^{-2} h_j $ & Successful & 0.6 & $ 8^\circ $ & 100 & $ 7.5\times 10^8~{\rm cm} $ & 0.2 & 3.0 & 500  \\
			$ \CW $ & $ 10^{-2} h_j $ & Choked & 0.6 & $ 8^\circ $ & 100 & $ 7.5\times 10^8~{\rm cm} $ & 0.2 & 0.225 & 500  \\\hline
			$ \HM $ & $ \sim 25 $ & Successful & 0.5 & $ \sim 12^\circ $ & $ \sim 1 $ & $ r_{h} $ & $\sim 1 $ & 0.4 & $\sim 100 $  \\
            
			\hline
			
		\end{tabular}
		\hfill\break
		
		\caption{Parameters of the different models. In the hydrodynamic and weakly magnetized jets the parameters are set as boundary conditions whereas in the highly magnetized jets the jet is launched self-consistently and the shown parameters are average values as found in the simulation. Shown are the initial jet magnetization $ \sigma_0 $; the specific jet enthalpy $ h_j $; the time delay from the merger to the jet launching, $ t_d $, during which the ejecta expanded homologously; the jet opening angle upon injection, $ \theta_j $; the jet injection height in the simulation, $ z_{\rm{beg}} $, where $ r_h = r_g(1+ \sqrt{1 - a^2}) $ is the event horizon; the jet total luminosity, $ L_j $; the engine working time, $ t_j $; the terminal Lorentz factor, if no mixing takes place, $ \Gamma_\infty $.
		}
		\label{tab:models}
	\end{table}

\section{Calibration of the shock breakout energy}\label{app:shock_breakout}

Once the jet-cocoon forward shock breaks out from the ejecta into an optically thin medium, it becomes collisionless, emit $ \gamma $-ray photons, and accelerates CR which interact with the photons in the breakout layer to generate neutrinos.
The shock breakout of the mildly relativistic cocoon is a leading candidate for the origin of the $ \gamma $-ray signal in GW170817, as it was found to be consistent with all the $ \gamma- $ray signal observables \citep{Gottlieb2018b}. Therefore, we use the properties of GRB170817A to estimate the breakout layer characteristics, assuming that cocoon shock breakout is indeed the physical mechanism behind the $ \gamma $-ray production and that GW170817 is a typical BNS. The $ \gamma $-ray isotropic equivalent energy of GRB170817A is $ E_{\rm iso,{\rm GRB17}} \approx 5 \times 10^{46} $ erg \citep{Goldstein2017}, and the breakout layer was at $ R_{\rm bo} \approx 2\times 10^{11}$ cm \citep{Gottlieb2018b}.
To find the energy and velocity angular structures of the jet-cocoon system, we utilize RMHD simulations of both successful and choked jets. For the successful jet we average the breakout properties of two high resolution simulations: $ II $ in \citet{Gottlieb2020c} (our model $ \SUa $), and $ S_b $ in \citet{Gottlieb2021a} (our model $ \SUb $), and for the uncollimated choked jet we use simulation $ \CU $ (see Appendix \ref{app:simulations} for the initial parameters of the models).

The energy originates in the range of angles between the minimal angle in which the Lorentz factor satisfies $ \Gamma^{-1} > \theta_{\rm obs}- \theta $, and $ \sim \theta_{\rm obs}$.
We calibrate the energy in this range in the simulations with the observed energy, and then calculate the energy observed by observers at other viewing angles.
Our simulations suggest that the width of $ \tau $ of a few in the merger frame is of a few percent of the shock radius, such that the width in the shock frame is $ \sim 10^{10}\Gamma^{-1} $ cm. in agreement with analytic estimates \citep[e.g.,][]{Nakar2012}.
At $ R_{\rm bo} $ the magnetic field becomes subdominant. By assuming adiabatic expansion of our weakly magnetized jet simulation, we find that the mass density is $ 3\times 10^{-8}~{\rm g~cm^{-3}} $, and the magnetic field is $ B(R_{\rm bo}) \approx 3\times 10^6 $ G. The latter is also obtained by applying the equipartition parameter $ \epsilon_{B} = 10^{-3} $ on the shock.

Finally, we estimate the $ \gamma $-ray photon spectrum to be composed of two phases of equal amount of energy. First, energy from the breakout layer is released without expansion of the breakout layer or thermalization of the photons, thus we use a Wien spectrum with a rest-frame peak at 50 keV, as appropriate for mildly relativistic radiation-mediated shocks. Over time emission from deeper layers is leaking and there is more time for the photons to thermalize and cool to a blackbody spectrum around 1 keV. For our accelerated CR energies, interaction with both $ \gamma $-ray and 1 keV X-ray photons can yield neutrinos.

\section{Neutrinos from choked jets}\label{app:choked}

Choked jets may produce neutrinos from subphotospheric shocks before or after they are choked. Here we show that after the jet is choked, cocoon shock breakout is the only neutrino production site (\S\ref{sec:successful}), and no neutrino emission is expected while the cocoon is still inside the ejecta.
The reason lies in heavy mixing of jet and cocoon material once the jet engine is shut off \citep{Gottlieb2021b}. The heavy mixing reduces the magnetization, thereby disfavoring the formation of collisionless shocks.
A relativistic jet can be choked due to a variety of reasons. Generally, choked jets can be sub-categorized to collimated and uncollimated choked jets. In case of the former, the jet engine is terminated before the jet finds its way outside the ejecta.

\begin{figure}[!h]
		\centering
		\includegraphics[scale=0.33]{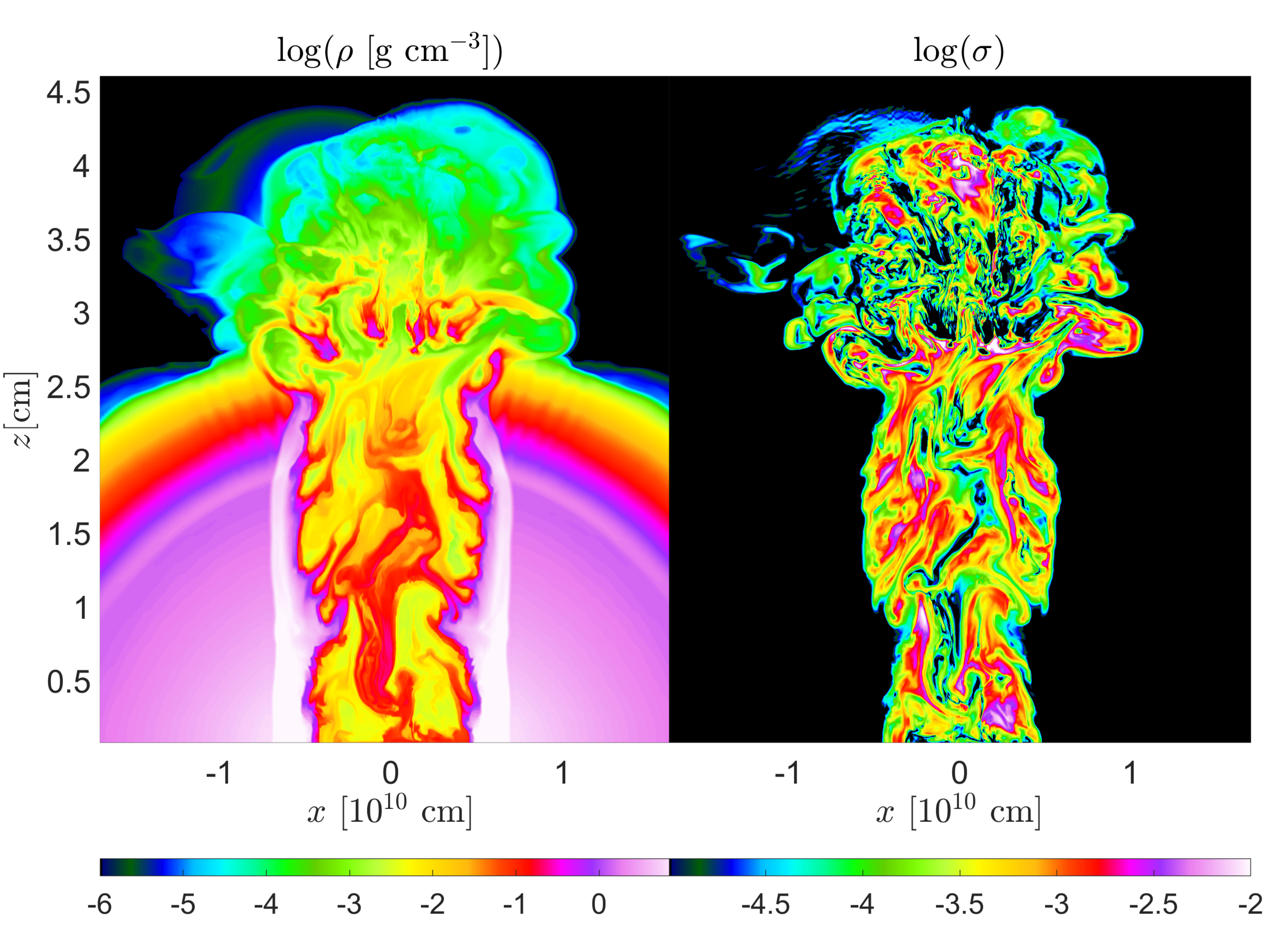}
		\caption[]{Logarithmic mass density (left) and magnetization (right) meridional maps of a jet that is choked upon breakout from the ejecta.
		}
		\label{fig:magnetic_choked_map}
\end{figure}

When the injected jet is uncollimated, the jet cannot punch through the dense medium, even if the central engine operates for a typical GRB duration or longer. In that case, the angular distribution of the outflow is more isotropic.
A collimated jet that is choked just before breakout has less time to mix with the cocoon material so it can keep its magnetization high, compared to a jet that is choked early-on or an uncollimated jet.
Thus, we perform a numerical simulation of such an optimal case (model $ \CW $), by fine tuning the collimated jet to be choked just upon breakout, we demonstrate that even under such conditions the mixing in the jet is too high to allow the formation of a collisionless subshock.
Fig. \ref{fig:magnetic_choked_map} depicts $ \hat{x}-\hat{z} $ meridional plane maps of the choked jet upon breakout. The jet cavity is filled with baryonic matter while the magnetized cocoon undergoes mixing with the surrounding unmagnetized ejecta such that its magnetization drops. With characteristic $ \sigma \sim 10^{-2.5} $ in the cocoon, it is unable to form collisionless subshocks in the outflow.
We conclude that even the most optimistic choked jet model, particle acceleration is inefficient. If the jet is choked earlier or is uncollimated in the first place, the resulting cocoon would maintain an even lower magnetization.

\end{document}